\documentclass[preprint,nofootinbib,amsmath,amssymb]{revtex4}

\usepackage[dvips]{graphicx}

\begin{document}
\title{Inhomogeneity-Induced Cosmic Acceleration in a Dust Universe}

\author{Chia-Hsun~Chuang}
\email{r92222016@ntu.edu.tw} %
\altaffiliation[\vspace{-1em} \\ \hspace*{0.1em} Now at ]{Homer
L.\ Dodge Department of Physics and Astronomy, Univ.\ of Oklahoma, \vspace{-1em} \\
\hspace*{0.1em} 440 W Brooks St., Norman, OK 73019, U.S.A.} %
\author{Je-An~Gu}
\email{jagu@phys.cts.nthu.edu.tw} %
\altaffiliation[\vspace{-1em} \\ \hspace*{0.1em} Now at ]{Physics
Division, National Center for Theoretical Sciences, P.O.\ Box 2-131, Hsinchu, Taiwan}%
%
\author{W-Y.~P.~Hwang}
\email{wyhwang@phys.ntu.edu.tw} %
\affiliation{Department of Physics, National Taiwan University, Taipei 10617, Taiwan, R.O.C.} %

\begin{abstract}
It is the common consensus that the expansion of a universe always
slows down if the gravity provided by the energy sources therein
is attractive and accordingly one needs to invoke dark energy as a
source of anti-gravity for understanding the cosmic acceleration.
To examine this point we find counter-examples for a spherically
symmetric dust fluid described by the Lemaitre-Tolman-Bondi
solution without singularity. Thus, the validity of this naive
consensus is indeed doubtful and the effects of inhomogeneities
should be restudied. These counter-intuitive examples open a new
perspective on the understanding of the evolution of our universe.
\end{abstract}

\maketitle


\section{Introduction\label{intro}}
In 1998, via the distance measurements of type Ia supernovae, it
was discovered that the expansion of the universe is accelerating
\cite{Perlmutter:1999np,Riess:1998cb}. The accelerating expansion
of the present universe was reinforced recently by the updated
supernova data
\cite{Tonry:2003zg,Knop:2003iy,Barris:2003dq,Riess:2004nr,Riess:2006fw}
and the WMAP measurement \cite{WMAP:2008} of cosmic microwave
background (CMB). Regarding the cosmic evolution, it is the common
consensus that normal matter (such as protons, neutrons,
electrons, etc.) can only provide attractive gravity and therefore
should always slow down the cosmic expansion, i.e.,
\begin{equation}
\textrm{Normal Matter} \, \Rightarrow \, \textrm{Attractive
Gravity} \, \Rightarrow \, \textrm{Deceleration} \, . %
\label{consensus}
\end{equation}
Thus, to explain this surprising, mysterious phenomenon of the
accelerating expansion, many people rely on exotic energy sources,
as generally called ``dark energy'', which provide significant
negative pressure and accordingly anti-gravity (repulsive
gravity).

The above conclusion about the existence of the cosmic
acceleration and the necessity of introducing dark energy is based
on a simplified cosmological model, the
Friedmann-Lemaitre-Robertson-Walker (FLRW) model, and indeed could
be model-dependent. In the FLRW model the universe is assumed to
be homogeneous and isotropic (i.e.\ the cosmological principle)
and accordingly the Robertson-Walker (RW) metric is invoked in the
Einstein equations that describe 
the evolution of the universe. Meanwhile the energy-momentum
tensor in the right-hand side of the Einstein equations is
regarded to truly reflect the real energy distribution (averaged
in space) of the universe. Nevertheless, so far there is no
convincing proof validating this simplification.

Apparently, our universe at present is not homogeneous at small
scales. The cosmological principle is roughly realized only at
very large scales. To take advantage of the cosmological principle
and invoke the RW metric, it is necessary to perform spatial
averaging over large scales, along with which the form of the
Einstein equations in general should change because of the
non-linearity of the Einstein equations \cite{Ellis:1984}. That
is, (i) when invoking the RW metric and the real energy sources 
of our universe in the Einstein equations for describing the
cosmic evolution, the left-hand side (the geometry part) of the
Einstein equations should be modified, or, from another point of
view, (ii) if one insists to use the Einstein tensor corresponding
to the RW metric in the left-hand side of the Einstein equations
(or, equivalently, moving the above mentioned modification in the
geometry part to the right-hand matter-energy part), there should
appear new, effective energy sources coming from geometry (which
is certainly ``dark'') and consequently the energy-momentum tensor
in the right-hand side does not truly correspond to the real
energy distribution of our universe. Thus, generally speaking, it
is doubtful to employ the Einstein equations to describe
long-time, large-scale phenomena, such as the evolution
of our universe, while the spatially averaged 
energy-momentum tensor and the spatially averaged metric tensor
are used therein. (For more discussions about the validity and the
problems of the FLRW cosmology, see \cite{Gu:2006}.)

Instead of invoking exotic energy sources or unconfirmed physics,
it is suggested
\cite{Buchert:1999mc,Schwarz:2002ba,Bene:2003fz,Rasanen:2003fy,Rasanen:2004js,Kolb:2004am,Barausse:2005nf,Kolb:2005me,Notari:2005xk,Rasanen:2005zy,Kolb:2005da,Mansouri:2005rf}
that the cosmic acceleration might originate from the violation of
the cosmological principle, homogeneity and isotropy, i.e., the
acceleration might be induced by the inhomogeneities of the universe. %
The possible change of the deceleration parameter in an
inhomogeneous universe has been pointed out in Refs.\
\onlinecite{Partovi:1982cg} and \onlinecite{Mashhoon:1983}. %
In particular, it has been shown that the luminosity
distance-redshift relation indicated by the supernova data can be
reproduced in an inhomogeneous cosmological model without
introducing dark energy
\cite{Mustapha:1998jb,Celerier:1999hp,Celerier:2000ew,Tomita:2000jj,Iguchi:2001sq,Alnes:2005rw,Apostolopoulos:2006eg,Garfinkle:2006sb,Biswas:2006ub,Enqvist:2006cg}. %


The current situation of our knowledge (what we know and what we
do not know well) about observations, cosmic acceleration and
inhomogeneities is as follows.\vspace{0.7em}

\noindent --- Known --- \\ 
$\bullet$\hspace{0.2em} \textit{Based on 
the FLRW cosmology, the current observational results indicate the
existence of the cosmic acceleration.} (That is, by invoking the
FLRW cosmological model to interpret the observational data, one
would conclude that the expansion of the universe is accelerating
in the recent epoch.) \vspace{0.7em}

\noindent --- Doubts --- \\ 
(1) \textit{Is the FLRW cosmology a good approximation?}\\
(2) \textit{Do the current observational results indicate the
existence of cosmic acceleration for the real universe with
complicated energy distribution?} (Even the definition of
accelerating expansion is an issue for our complicated universe.
This issue will be discussed in Sec.\ \ref{def}.)\\
(3) \textit{Can the inhomogeneities of our universe explain the
observational results?}\\
(4) \textit{Can the inhomogeneities of our universe generate
accelerating expansion?}\vspace{0.7em}

These four doubts are far from being fully answered. This is due
to the difficulties from the complexity of the real energy
distribution and the non-linearity of the Einstein equations.
Because of these difficulties, instead of dealing with the full
Einstein equations describing the real complicated universe,
usually people employ the following two approaches to sketch the
possible answers to the above doubts.%
\vspace{0.2em}

\noindent %
(I) \textit{Taking perturbative approach with a simple background}
(such as the homogeneous and isotropic RW background).\\ %
(II) \textit{Invoking exact solutions of the Einstein equations
describing an inhomogeneous universe}.\vspace{0.5em}

\noindent %
In addition to these two approaches, there are also
non-perturbative studies without invoking exact solutions, e.g.,
see Refs.\
\onlinecite{Sicka:1999cb,Buchert:1999pq,Rasanen:2008it}.

Via Approach (I) with a homogenous and isotropic background, the
positive answer to Doubt (1) is supported in Refs.\
\onlinecite{Siegel:2005xu} and \onlinecite{Ishibashi:2005sj}, and
the negative answer to Doubts (3) and (4) supported in Refs.\
\onlinecite{Siegel:2005xu,Ishibashi:2005sj,Kasai:2006bt,Vanderveld:2007cq}.
%
Nevertheless, the reliability of the perturbative approach in (I)
is doubtful for investigating the late-time cosmic evolution, for
which not only energy distribution, but also curvature, may not be
described with perturbations \cite{Kolb:2005da,Rasanen:2006kp}.
Furthermore, the arguments of Refs.\
\onlinecite{Siegel:2005xu,Ishibashi:2005sj,Kasai:2006bt,Vanderveld:2007cq}
have been countered in Refs.\
\onlinecite{Kolb:2005da,Kolb:2005ze,Rasanen:2006kp,Rasanen:2008it}.

The drawback of Approach (II) is that the exact solution invoked
may be very different from the real situation of our universe.
From some angle, instead of Doubts (3)
and (4), Approach (II) is to answer a more general question:%
\vspace{0.2em}

\noindent %
$\bullet$\hspace{0.2em} \textit{Can inhomogeneities
explain observational results and generate accelerating expansion?}%
\vspace{0.2em}

\noindent %
Or, more precisely,\vspace{0.3em}

\noindent (3$^{\prime}$) \textit{Does there exist a universe
(maybe very different from ours) in which the
inhomogeneities can explain the observational results?}\\
(4$^{\prime}$) \textit{Does there exist a universe in which the
inhomogeneities can drive the expansion to
accelerate?}\vspace{0.5em}

\noindent %
Via Approach (II), for Doubt (3$^{\prime}$) it has been shown in
Refs.\
\onlinecite{Mustapha:1998jb,Celerier:1999hp,Celerier:2000ew,Tomita:2000jj,Iguchi:2001sq,Alnes:2005rw,Apostolopoulos:2006eg,Garfinkle:2006sb,Biswas:2006ub,Enqvist:2006cg}
that the supernova data can be explained by invoking the
Lemaitre-Tolman-Bondi solution
\cite{Lemaitre:1933qe,Tolman:1934za,Bondi:1947av} that describes a
spherically symmetric (but inhomogeneous) dust fluid.

In the present work, we take Approach (II) for investigating the
general possibility of generating accelerating expansion via
inhomogeneities, i.e.\ Doubt (4$^{\prime}$). 
Against the common intuition in Eq.\ (\ref{consensus}) we find
examples of accelerating expansion in the case of a spherically
symmetric dust fluid described by the Lemaitre-Tolman-Bondi (LTB)
solution, thereby giving support to the positive answer to Doubt
(4$^{\prime}$). %
In addition to our examples, Kai et al.\ \cite{Kai:2006ws} also
found acceleration examples based on the LTB solution, where
there exists singularity around the center of the spherically
symmetric system during the accelerating epoch. In contrast, in
our examples the system is smooth everywhere and no singularity is
involved. (For more comparison, see Sec.\ \ref{dom}.) %
%
The possibility of the accelerating expansion in the LTB model was
also pointed out by Paranjape and Singh in Ref.\
\onlinecite{Paranjape:2006cd}, in which numerical models
exhibiting acceleration were constructed by an approximation where
the contribution of 3-curvature dominates over the matter density.

%
There also existed acceleration examples for a system consisting
of two or more regions
\cite{Nambu:2005zn,Rasanen:2006kp,Rasanen:2006zw,Rasanen:2008it,Wiltshire:2007fg,Paranjape:2008ai}.
In these acceleration examples, connecting the separate regions
smoothly and the effect of the junction between these regions are
the essential issues yet to be seriously explored. %
We particularly note that it needs much caution to connect two (or
more) regions. In many cases the effect from the junction between
separate regions is significant and should not be ignored. %
A similar doubt was also raised by Paranjape and Singh in Ref.\
\onlinecite{Paranjape:2008ai}. %
As a demonstration of how things may go wrong when the connection
or the junction is not appropriately taken care, in Appendix
\ref{app:NT_example} we investigate in detail the acceleration
examples studied by Nambu and Tanimoto in Ref.\
\onlinecite{Nambu:2005zn}, and show that actually there should be
no acceleration in those cases when we properly connect the
separate regions and seriously take the
effect of the junction into account. %
%

The acceleration examples we find could be far away from the real
situation of our universe.
We are not proposing to employ these mathematical examples or the
LTB solution to describe our universe. These acceleration
examples, which provide positive answer to Doubt (4$^{\prime}$),
are to demonstrate how inhomogeneities can drive the expansion to
accelerate, thereby showing how our intuition about the interplay
of gravity and the cosmic evolution may go wrong [i.e.\ against
the common intuition in Eq.\ (\ref{consensus})]. Accordingly, the
effects of inhomogeneities on the evolution of the universe should
be carefully restudied.

This paper is organized as follows. In Sec.\ \ref{def}, a tricky
issue of the definition of acceleration is discussed and two
definitions to be utilized for searching acceleration examples are
introduced. In Sec.\ \ref{ltb}, the LTB solution is described. In
Sec.\ \ref{rad} and Sec.\ \ref{dom}, we present the examples of
the accelerating expansion corresponding to these two definitions
of acceleration, respectively. A summary and discussions follow in
Sec.\ \ref{con}.

Throughout the present paper we will use the units where $c = 8
\pi G = 1$. We note that in this unit system there is still one
unit unspecified, and, as a result, the value of one of the
dimensionful quantities can be arbitrarily set. For example, in
the acceleration examples we will present, the physical size of
the spherical region under consideration can be any length long
(such as $1\,$fermi, $1\,$cm, $1\,$pc, $1\,$Mpc, Hubble length
$H_0^{-1}$, etc.). Once the value of one of the dimensionful
quantities (which are independent of $c$ and $G$) is settled, all
the units for the dimensionful quantities studied in the present
work are specified.


\section{Definitions of Acceleration\label{def}}

In cosmology, expansion and acceleration of the universe are a
subject associated with the evolution of the space (in size), and
should have nothing to do with the particle motion relative to the
space. How to define the speed and the acceleration purely
corresponding to the evolution of the space, meanwhile avoiding
confusion and interference from the particle motion relative to
the space? This is a tricky issue for a real universe with
complicated energy distribution \cite{Gu:2006}.

To characterize the evolution status of the space, usually one
needs to invoke two quantities: a length (distance) quantity $L$
and a time quantity $t$, with which $\dot{L}$ and $\ddot{L}$
(where the over-dot denotes the time derivative) present the speed
and the acceleration of the expansion, respectively. The tricky
issue mentioned above then corresponds to the problem of making
the choice of these two quantities. With different choices one has
different definitions. It will be puzzling if the description of
the space evolution (expansion and acceleration) status is not
universal but depends on the definition one invokes.%
%

If it is inevitable to make the choice which the description of
the evolution status of the universe is based on and maybe
sensitively depends on, what is the reasonable physical choice?
The very guiding principle for making the choice is that we are
quantifying simply the evolution of the space and we need to make
sure that the definition we invoke does not involve particle
motion relative to the space nor something fake stemming from an
inappropriate frame choice.

In this section we will introduce two definitions, as to be called
``\textit{line acceleration}'' and ``\textit{domain
acceleration}'', respectively involving two length quantities: (i)
the distance between two points in space and (ii) the size of a
domain in space. For these two kinds of lengths the above tricky
issue relates to the choice of two points for the line
acceleration and the choice of the domain (i.e.\ its boundary) for
the domain acceleration. With different choices (corresponding to
different definitions) one might obtain very different conclusions
about the acceleration/deceleration status.

In general, with a certain choice of frame (coordinate system),
the two spatial points and the boundary mentioned above can be
fixed in the spatial coordinate space. With an improper choice of
these two points and the boundary, or, equivalently, with an
improper choice of frame (in which the spatial coordinates of the
two points and the boundary are fixed), the change of the length
quantity with time may be incapable of representing the true
evolution of the space, which could be mixed up with the evolution
of the frame. The contribution from the frame evolution to the
acceleration of the chosen length will be called ``frame
acceleration'' in the present paper.

To construct suitable definitions of acceleration for these two
kinds of lengths, for simplicity we consider a universe consisting
of freely-moving particles (i.e., moving along geodesics), among
which there is no interaction other than gravity. In this case,
for the line acceleration a simple reasonable length quantity to
choose is the distance between two (freely-moving) particles. In
contrast, an apparently improper definition which could lead to
fake frame acceleration is to invoke the distance between two
points which move ``outward'' relative the particles in between,
i.e., accordingly, there are more and more particles between these
two chosen spatial points.

As to the domain acceleration, a reasonable choice is the size of
a spatial domain in which the number of particles is constant in
time. In contrast, an apparently improper definition which could
lead to fake frame acceleration is to invoke a domain whose
boundary moves outward relative the particles therein, i.e.,
accordingly, there are more and more particles within this domain.

For implementing these two acceleration definitions, meanwhile
satisfying the above mentioned requirements for avoiding the
confusion from particle motion and fake frame acceleration, it is
particularly beneficial to use the synchronous 
gauge. %
Note that in a dust universe the synchronous gauge can be chosen
if and only if the vorticity vanishes. %
In the synchronous gauge the line element is as follows.
\begin{equation}
ds^2=-dt^2 + h_{ij}({\bf x},t) dx^i dx^j  \, ,
\end{equation}
where $t$ is the cosmic time. %
In this gauge, the cosmic time $t$ is simple and universal to
choose in defining acceleration. Regarding the length quantity, in
this gauge the above mentioned requirements are easy to meet
because a point fixed in the spatial coordinate space is a
geodesic. In particular, for a dust fluid with the energy-momentum
tensor
\begin{equation}
T^{\mu \nu}=\rho u^\mu u^\nu \, , \quad u^\mu=(1,0,0,0) \, ,
\end{equation}
the distance between two fixed points and the size of a domain
with its boundary fixed in this coordinate space are simple and
direct choices satisfying the requirements. In the following we
will focus on this simple case of a dust fluid described in the
synchronous gauge, and introduce two definitions of acceleration
involving these two length quantities respectively.

\subsection{Line Acceleration}
Regarding two points in space with the proper distance $L(t)$
between them at time $t$, it is reasonable to use $\dot{L}(t)$ and
$\ddot{L}(t)$ to characterize the expansion/collapse status and
the acceleration/deceleration status of the space in between. The
expansion rate and the deceleration parameter for the proper
distance $L(t)$ are defined in the usual way, as follows:
\begin{eqnarray}
H_L & \equiv & \dot{L}/L \, , \\
q_L & \equiv & -\frac{\ddot{L}/L}{H_L^2} =
-\frac{\ddot{L}L}{\dot{L}^2} \, .
\end{eqnarray}
The condition $\{ H_L > 0 \, , \, q_L < 0 \}$ corresponds to the
accelerating expansion of the proper distance between these two
points in space, which is dubbed ``line acceleration'' in the
present paper. We have found examples of the line acceleration in
a dust universe, as to be shown in Sec.\ \ref{rad}.

\subsection{Domain Acceleration}
Another definition of acceleration, as dubbed ``domain
acceleration'' in the present paper, has been widely used in the
literature
\cite{Buchert:1999er,Sicka:1999cb,Buchert:1999mc,Buchert:1999pq,Palle:2002zf,Kolb:2005da,Nambu:2005zn,Rasanen:2006zw,Rasanen:2006kp,Wiltshire:2007fg,Paranjape:2008ai,Rasanen:2008it}.
It is for a spatial domain $D$ with a finite volume
\begin{equation}\label{vd}
V_D \equiv \int_D\, \sqrt{h}\,d^3\!x ,
\end{equation}
where $h$ is the determinant of the spatial metric tensor
$h_{ij}$. Invoking the length scale of the domain,
\begin{equation}
L_D \equiv V_D^{1/3} \, , %
\label{domain size}
\end{equation}
one can define the expansion rate and the deceleration parameter
of the domain in the usual way, as follows:
\begin{eqnarray}
H_D & \equiv & \dot{L}_D/L_D \, , \\
q_D & \equiv & -\frac{\ddot{L}_D/L_D}{H_D^2} = %
-\frac{\ddot{L}_D L_D}{\dot{L}_D^2} \, . \label{qD def}
\end{eqnarray}
The condition $\{ H_D > 0 \, , \, q_D < 0 \}$ corresponds to the
accelerating expansion of the domain (in size), i.e., domain
acceleration.

As shown in
\cite{Flanagan:2005dk,Hirata:2005ei,Giovannini:2005sy,Alnes:2005nq},
for an infinitesimal domain in a dust universe %
without vorticity %
the deceleration parameter $q_D$ is always positive, i.e.,
corresponding to local deceleration. %
Nevertheless, the non-local deceleration/acceleration status of a
domain with a nonzero finite volume (in particular, the
observational universe of the Hubble size) may be very different
\cite{Kolb:2005da}. %
So far there is no no-go theorem excluding the possibility of
negative $q_D$. On the contrary, we have found examples of the
domain acceleration in a dust universe, as to be shown in Sec.\
\ref{dom}.

We note that, in the special case of a homogeneous and isotropic
universe described by the RW metric with the scale factor $a(t)$,
the expansion rates and the deceleration parameters defined above
are the same as those in the standard cosmology:
\begin{eqnarray}
H & \equiv & \dot{a}/a \, , \\
q & \equiv & -\frac{\ddot{a}/a}{H^2} =
-\frac{\ddot{a}a}{\dot{a}^2} \, .
\end{eqnarray}

In the next section we will introduce the LTB solution, based on
which we find examples of acceleration. %
(Reminder: the units where $c = 8 \pi G = 1$ will be employed.)


\section{Lemaitre-Tolman-Bondi (LTB) Solution\label{ltb}}
The LTB solution \cite{Lemaitre:1933qe,Tolman:1934za,Bondi:1947av}
is an exact solution of the Einstein equations for a spherically
symmetric dust fluid. The metric is given by
\begin{eqnarray}
\label{eq1} %
ds^2 = -dt^2  + \frac{\left(R,_{r}\right)^2 dr^2}{1 + 2\,E(r)}+R^2
d\Omega^2 \, ,
\end{eqnarray}
where $R$ is a function of the time coordinate $t$ and the radial
coordinate $r$, $E(r)$ is an arbitrary function of $r$, and
$R,_{r}$ denotes the partial derivative of $R$ with respect to
$r$. With this metric the Einstein equations can be reduced to two
equations:
\begin{eqnarray}
\label{eq2} \left({\frac{\dot{R}}{R}}\right)^2&=&\frac{2
E(r)}{R^2}+\frac{2M(r)}{R^3} \, , \\
\label{eq3} \rho(t,r)&=&\frac{2M'(r)}{R^2 R,_{r}} \, ,
\end{eqnarray}
where $M(r)$ is an arbitrary function of $r$ and the over-dot
denotes the partial derivative with respect to $t$. The solution
of Eq.\ (\ref{eq2}) can be written parametrically by using a
variable $\eta=\int dt/R \,$, as follows.
\begin{eqnarray}
\label{eq4} R(\eta ,r) &=& \frac{M(r)}{- 2 E(r)}
     \left[ 1 - \cos \left(\sqrt{-2 E(r)} \eta \right) \right] \, ,\\
\label{eq5} t(\eta ,r) &=& \frac{M(r)}{- 2 E(r)}
     \left[ \eta -\frac{1}{\sqrt{-2 E(r)} } \sin \left(\sqrt{-2 E(r)}
     \eta \right) \right] + t_{b}(r) \, ,
\end{eqnarray}
where $t_{b}(r)$ is an arbitrary function of $r$. Summarily, there
are one dynamical field, $R(t,r)$, and three arbitrary functions,
$E(r)$, $M(r)$ and $t_b(r)$. For a given set of the three
functions $\{ E(r),M(r),t_b(r) \}$, we have a solution $R(t,r)$
specified by Eqs.\ (\ref{eq4}) and (\ref{eq5}).

Regarding the behavior of the dynamical field and the three
functions introduced above, it is reasonable to consider the
requirements that there is no hole and no singularity in space and
the energy density is non-negative and finite. \\
(1) For no hole at the center (i.e.\ the area of the spherical
surface at $r=r_0$ goes to zero when $r_0$ goes to zero), $R(t,r=0)=0$. \\
(2) For no singularity, we consider $R,_{r}(t,r) \neq 0$. \\
(3) For a non-negative and finite energy density, $0 \leq \rho(t,r) < \infty$. \\
According to these requirements, the dynamical field and the
functions in the LTB solution should satisfy the following
restrictions:
\begin{eqnarray}
r=0 & : & \; R , \ddot{R} , M , E = 0 \, , \\
r>0 & : & \; RR,_{r} > 0 \; , \; MM' \geq 0 \, .
\end{eqnarray}
Without losing generality, in our study and in the remaining of
the present paper we choose
\begin{equation} \label{eq:sign choice}
R , R,_{r} , M , M' \geq 0 \, .
\end{equation}
From the relation
\begin{equation}
\ddot{R} = - \frac{M(r)}{R^2} \, ,
\end{equation}
as derived from the Einstein equation (\ref{eq2}), the sign choice
in Eq.\ (\ref{eq:sign choice}) implies
\begin{equation} \label{eq:ddotR < 0}
\ddot{R} \leq 0 \, .
\end{equation}
For more details about the restrictions on the LTB solution and
the corresponding features of the dynamical field and the
functions presented above, see Appendix \ref{app:ltb_constraint}.
Note that the necessary and sufficient conditions of no shell
crossing in a period of time are presented by Hellaby and Lake in
Ref.\ \onlinecite{Hellaby:1985zz}.
In the present paper we consider the absence of shell crossing at
some time, and accordingly the above-mentioned conditions are
necessary conditions.

By introducing the following variables
\begin{equation}
 a(t,r)=\frac{R(t,r)}{r},\quad k(r)=-\frac{2E(r)}{r^2},\quad
  \rho_0(r)=\frac{6M(r)}{r^3} \, ,
\end{equation}
the line element in Eq.\ (\ref{eq1}) and the Einstein equations
(\ref{eq2}) and (\ref{eq3}) can be rewritten in a form more
similar to that of the RW metric:
\begin{equation}
\label{eq6} ds^2 =
-dt^2+a^2\left[\left(1+\frac{a,_{r}r}{a}\right)^2
    \frac{dr^2}{1-k(r)r^2}+r^2d\Omega_2^2\right] \, ,
\end{equation}
\begin{eqnarray}
\label{eq7} %
\left(\frac{\dot{a}}{a}\right)^2 &=&
-\frac{k(r)}{a^2}+\frac{\rho_0(r)}{3a^3} \, ,\\
\label{eq:LTB rho 2} %
\rho(t,r) &=& \frac{(\rho_0 r^3)'}{3 a^2 r^2 (ar)_{, r}} \, .
\end{eqnarray}
The solution in Eqs.\ (\ref{eq4}) and (\ref{eq5}) becomes
\begin{eqnarray}
\label{LTB soln2 R} a(\tilde{\eta},r) &=& \frac{\rho_0(r)}{6k(r)}
     \left[ 1 - \cos \left( \sqrt{k(r)} \, \tilde{\eta} \right) \right] \, ,\\
\label{LTB soln2 t} t(\tilde{\eta},r) &=& \frac{\rho_0(r)}{6k(r)}
     \left[ \tilde{\eta} -\frac{1}{\sqrt{k(r)}} \sin
     \left(\sqrt{k(r)} \, \tilde{\eta} \right) \right] + t_{b}(r) \, ,
\end{eqnarray}
where $\tilde{\eta} \equiv \eta r = \int dt/a \,$.

Now the dynamical field describing the evolution of the space is
$a(t,r)$ and the three functions to be given for specifying a
solution $a(t,r)$ are $k(r)$, $\rho_0(r)$ and $t_b(r)$
[corresponding to $E(r)$, $M(r)$ and $t_b(r)$, respectively].
Apparently, when all the functions $a(t,r)$, $k(r)$, $\rho_0(r)$
and $t_b(r)$ have no dependence on the radial coordinate $r$, we
come back to the RW metric from Eq.\ (\ref{eq6}), the Friedmann
equation from Eq.\ (\ref{eq7}) and the formula realizing stress
energy conservation from Eq.\ (\ref{eq:LTB rho 2}). From the
comparison with the RW metric, we can get a rough picture of the
LTB metric: $a(t,r)$ plays the role of a spatially varying,
time-dependent scale factor describing the evolution of the space,
$k(r)$ corresponds to the spatial curvature, $\rho_0(r)$ relates
to the physical energy density $\rho(t,r)$, and $t_b(r_1)$ can be
regarded as the initial time of the big bang at $r=r_1$, i.e., the
space at $r=r_1$ starts to expand from singularity
[$a(t_b,r_1)=0$] at the time $t=t_b(r_1)$.

In search of examples of accelerating expansion in a dust universe
described by the LTB solution, we tried a variety of LTB solutions
corresponding to different choices of the functions $\{ k(r) ,
\rho_0(r) , t_b(r) \}$, and eventually found examples among these
tedious trials, as to be studied in the following two sections. We
note that there is redundancy in the choices of the functions $\{
E(r) , M(r) , t_b(r) \}$ or $\{ k(r) , \rho_0(r) , t_b(r) \}$. For
example, for a monotonically increasing function $M(r) \equiv
\rho_0(r) r^3$, one can choose, without losing generality,
\begin{equation}
\rho_0(r) = \mbox{constant} \, ,
\end{equation}
which is the choice invoked in our search for the acceleration
examples.


\section{Line Acceleration induced by Inhomogeneities\label{rad}}

Regarding the demonstration of how the acceleration can be induced
by inhomogeneity, naively we have better chance with larger
inhomogeneity which may induce more significant acceleration. In
the LTB solution, due to the spherical symmetry, the inhomogeneity
lies along the radial direction, while there is no inhomogeneity
and also no acceleration [as implied by Eq.\ (\ref{eq:ddotR < 0})]
in the angular directions. Accordingly the possible line
acceleration induced by inhomogeneity must involve the length
component in the radial direction. For simplicity, we focus on the
proper distance between the origin ($r=0$) and the point at
$r=r_L$ at time $t\,$:
\begin{equation}\label{eq:lr}
L_r(t) \equiv \int_0^{r_L} \sqrt{g_{rr}}dr \, ,
\end{equation}
where
\begin{equation} \label{eq:grr}
g_{rr} = \frac{\left(R,_{r}\right)^2}{1+2\,E(r)} =
\frac{\left(a+a,_{r}r\right)^2}{1-k(r)r^2} \, .
\end{equation}
The deceleration parameter corresponding to this radial proper
distance is
\begin{equation}
q_r \equiv -\frac{\ddot{L}_r L_r}{\dot{L}_r^2} \, .
\end{equation}
The sign of the deceleration parameter $q_r$ is determined by the
sign of $\ddot{L}_r$ or, more precisely, the integral of
$\partial_t^2\sqrt{g_{rr}}$ from the origin $r=0$ to $r=r_L$,
i.e., $\int_{0}^{r_L}\partial_{t}^{2}\sqrt{g_{rr}}dr$.

\begin{figure}[h]
  \center
  \includegraphics[width=0.65\linewidth,clip]{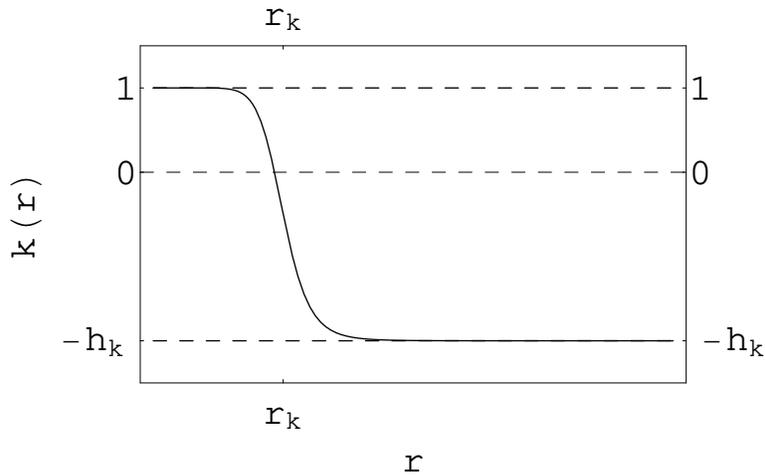}
  \caption{The plot of the function $k(r)$ invoked in the search for
  acceleration examples.}
  \label{fig:k(r)}
\end{figure}

In the LTB solution, inhomogeneity can be introduced by choosing
inhomogeneous functions for $k(r)$, $\rho_0(r)$, and $t_b(r)$. In
this section, for simplicity we introduce inhomogeneity through
only $k(r)$ by employing the following function:
\begin{equation}
k(r) = -\frac{(h_k+1) (r/r_k)^{n_k}}{1+(r/r_k)^{n_k}}+1 \, ,
\label{k function}
\end{equation}
while choosing
\begin{eqnarray}
\rho_0 (r) &=& \textrm{constant} \, , \label{const rho0 function} \\
t_b(r) &=& 0 \, . \label{trivial tb function}
\end{eqnarray}
The behavior of the function $k(r)$ in Eq.\ (\ref{k function}) is
illustrated in Fig.\ \ref{fig:k(r)}. For a large power $n_k$, this
$k(r)$ function mimics a step function with violent change around
$r=r_k$, accompanying which large inhomogeneity is introduced. We
note that when $k(r)=0=E(r)$ [with arbitrary $\rho_0(r)$ and
$t_b(r)$] there is no line acceleration (see Appendix
\ref{app:E=0}), and therefore the inhomogeneity in $k(r)$ seems to
play an essential role in generating accelerating expansion.

With the above choice of the three arbitrary functions in the LTB
solution, we have six free parameters to tune:
\begin{equation}
(t,r_L,\rho_0,r_k,n_k,h_k) \, . %
\label{radial parameters}
\end{equation}
In search of examples of the line acceleration in the radial
direction, we surveyed this six-dimensional parameter space and
did find examples eventually.%
\footnote{For $1-k(r)r^2$ to be positive, we consider two
different sufficient conditions, $\{ n_k > 0 , h_k > 1 , r_k < 1
\}$ and $\{ h_k > -1 , r \leq r_L \leq 1 \}$, and restrict our
search in the cases satisfying one of them.}
Table \ref{table:radial eg} presents one of the examples with
significant acceleration, $-q_r \cong 0.834 \sim \mathcal{O}(1)$.
We note that, as suggested by the observational data with the
analyses based on the FLRW cosmology, the deceleration parameter
of our present universe is of the order unity and is negative in
sign.

\tabcolsep=8pt
\begin{table}[h!]
\caption{One example of the line acceleration in the radial
direction.} \label{table:radial eg}
\begin{tabular}{|c|c|c|c|c|c||c|c|c|c|}\hline
$t$&$r_L$&$\rho_0$&$r_k$&$n_k$&$h_k$&$L_r$&$\dot{L}_r$&$\ddot{L}_r$&$q_r$  \\
\hline
$1$&$1$&$1$&$0.7$&$20$&$1$&$1.29$&$1.16$&$0.868$&$-0.834$  \\
\hline
\end{tabular}
\end{table}

\begin{figure}[h!]
  \center
  \includegraphics[width=0.6\linewidth,clip]{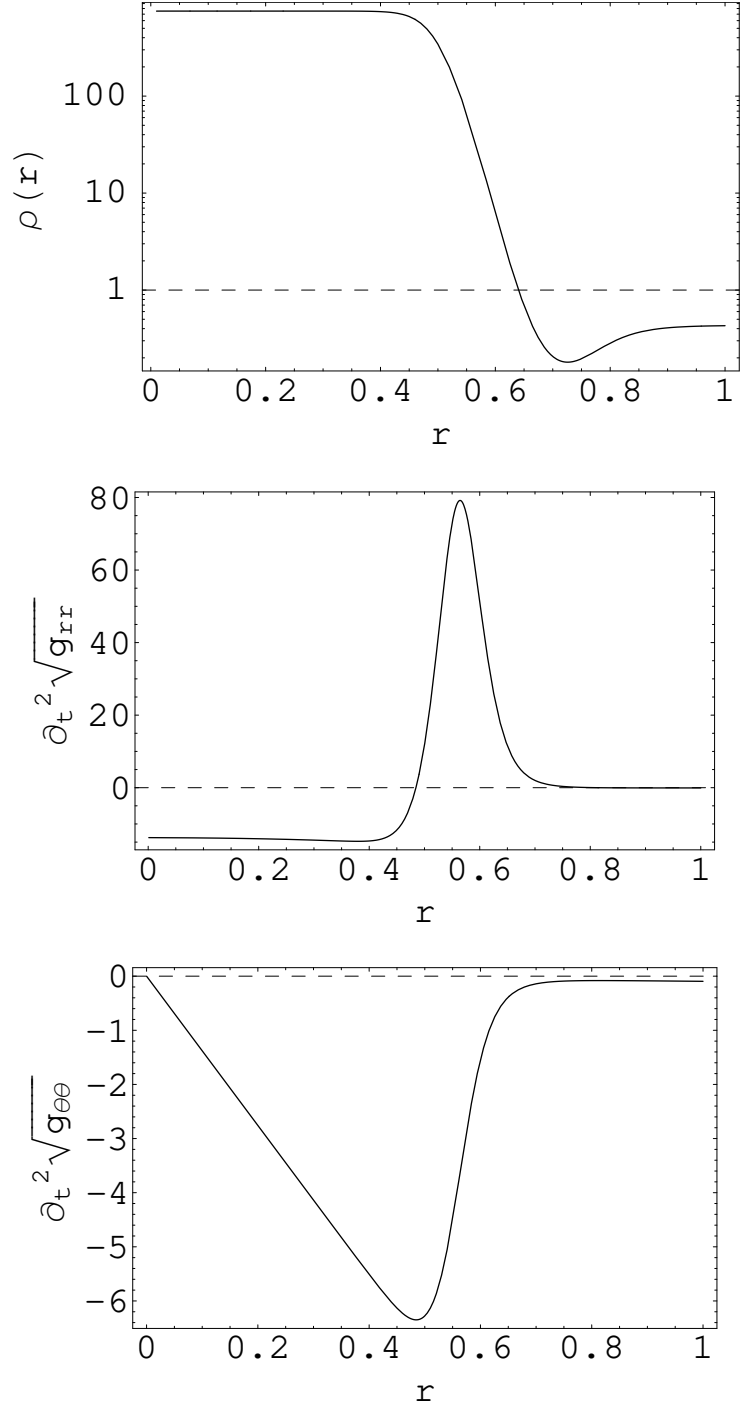}
  \caption{The plots of the physical energy density $\rho$ and the
  quantities, $\partial_t^2\sqrt{g_{rr}}$ and $\partial_t^2\sqrt{g_{\theta
  \theta}}$, which characterize the local acceleration/deceleration
  status in the radial and the angular direction, respectively,
  for the example in Table \ref{table:radial eg}.
  We note that when $r=r_L=1$, $\partial_t^2\sqrt{g_{rr}}=-0.066$
  and $\partial_t^2\sqrt{g_{\theta \theta}}=-0.095\,$.%
  }
  \label{fig:rad}
\end{figure}

The further details of the example in Table \ref{table:radial eg}
are illustrated in Fig.\ \ref{fig:rad}, which presents the energy
density distribution and the acceleration/deceleration status of
the radial and the angular line elements. As shown in this figure,
the spherically symmetric dust fluid consists of three regions:
two roughly homogeneous regions
--- the inner over-density region with positive $k(r)$ and smaller
$a(t,r)$ and the outer under-density region with negative $k(r)$
and larger $a(t,r)$ --- and one transition or junction region,
where the inhomogeneity locates, of those two homogeneous
regions.\footnote{Note that in the inner region around the origin
$r=0$ the energy density distribution is flat and therefore there
is no singularity and, moreover, no cusp behavior (or weak
singularity) in this acceleration example.} In the inhomogeneous
region with significantly changing energy density we have
acceleration in the radial line elements, i.e., $\partial_t^2
\sqrt{g_{rr}} > 0$, while in the homogeneous regions with smoothly
distributed energy density we have deceleration. This result
strongly supports the suggestion that inhomogeneity can induce
accelerating expansion. In addition, as expected there is always
no acceleration in the angular directions
along which everything is uniformly distributed.%
\footnote{For a proof, see Appendix \ref{app:ltb_constraint}. The
result is in Eq.\ (\ref{ltb_constraint11}).}

For a demonstration of the scales of the size and other quantities
of this system, in Table II  
we fix the only one unspecified unit by using the length unit:
$0.1$Mpc, $1$Mpc, $10$Mpc and $100$Mpc, and present the values of
several
dimensionful quantities, 
respectively. Among these quantities, $t$ corresponds to the time
under consideration, $L_r$ the size of the system, $\rho(r=0)$
roughly the energy density of the inner region and $\rho(r=r_L)$
roughly that of the outer region. %
As shown in this table, regarding the same example in Table
\ref{table:radial eg}, when the size of the system increases by
one order of magnitude, the time increases by one order and the
energy density decreases by two orders of magnitude. For this
example to be consistent with the situation of our present
universe, the time $t$ should be $\sim 10^{10}\,$years and the
energy density of the outer region should be similar to the
average energy density of the present universe, $\sim
10^{-29}\,$g/cm$^3$. One can see that these two conditions cannot
be simultaneously satisfied in this example, and therefore this
example by itself alone cannot describe the present universe.

\begin{table}[h!]  \label{table:rq-units}
\caption{Corresponding to different length units, the values of
several dimensionful quantities for the acceleration example in
Table \ref{table:radial eg} are presented.} %
\center
\begin{tabular}{|c|c|c|c|c|}\hline
Length unit & $L_r$(Mpc) & $t$(year)
& $\rho(r=0)$ (g/cm$^3$) & $\rho(r=r_L)$ (g/cm$^3$) 
\\
\hline 0.1 Mpc & 0.129 & 3.26E5 & 1.19E-29 & 6.80E-33 
\\
\hline 1 Mpc & 1.29 & 3.26E6 & 1.19E-31 & 6.80E-35 
\\
\hline 10 Mpc & 12.9 & 3.26E7 & 1.19E-33 & 6.80E-37 
\\
\hline 100 Mpc & 129 & 3.26E8 & 1.19E-35 & 6.80E-39 
\\
\hline
\end{tabular}
\end{table}

\begin{figure}[h!]
  \raggedright
  \includegraphics[width=0.45\linewidth,clip]{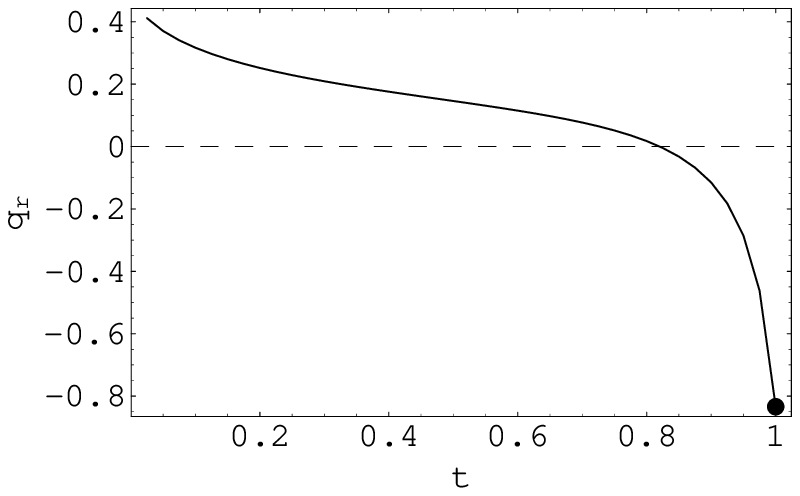}
  \includegraphics[width=0.45\linewidth,clip]{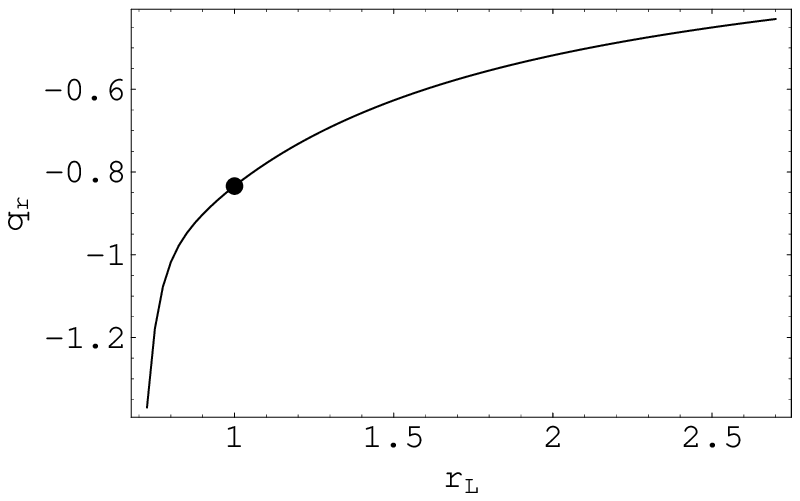}
  \includegraphics[width=0.45\linewidth,clip]{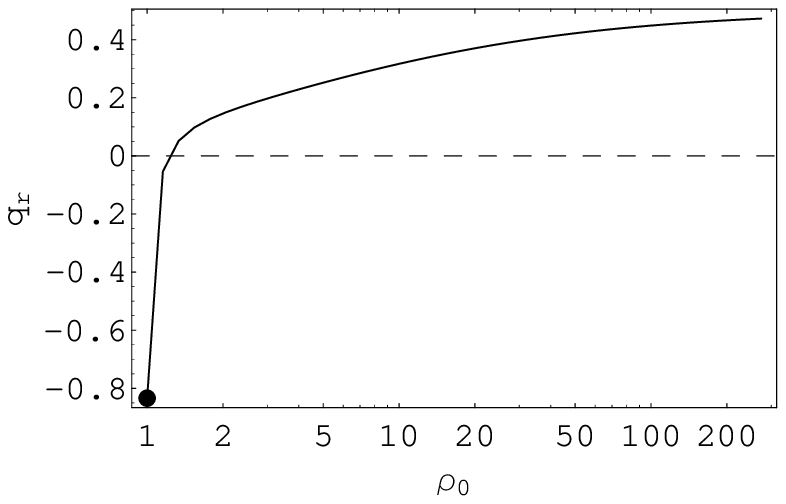}
  \includegraphics[width=0.45\linewidth,clip]{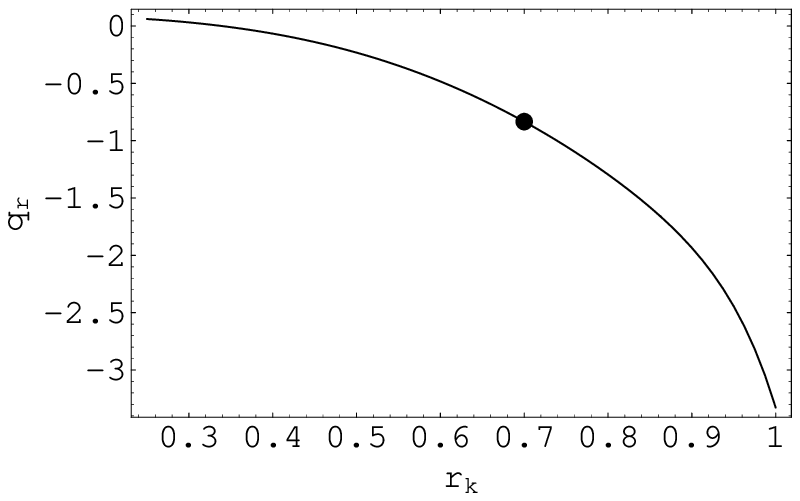}
  \includegraphics[width=0.45\linewidth,clip]{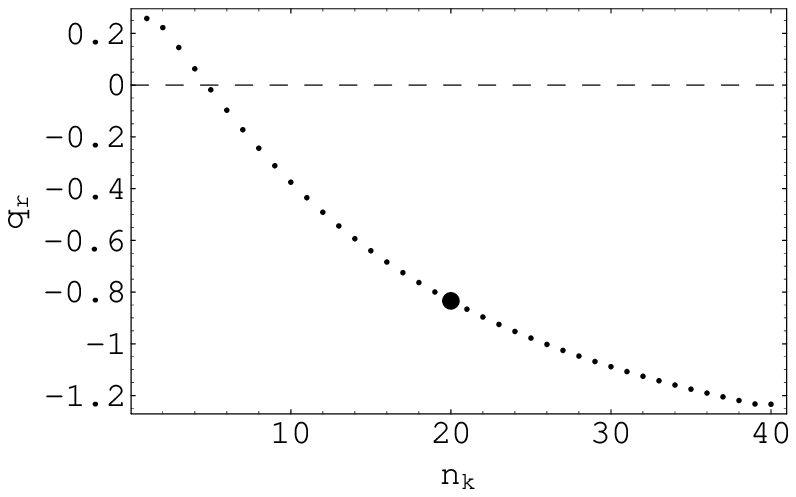}
  \includegraphics[width=0.45\linewidth,clip]{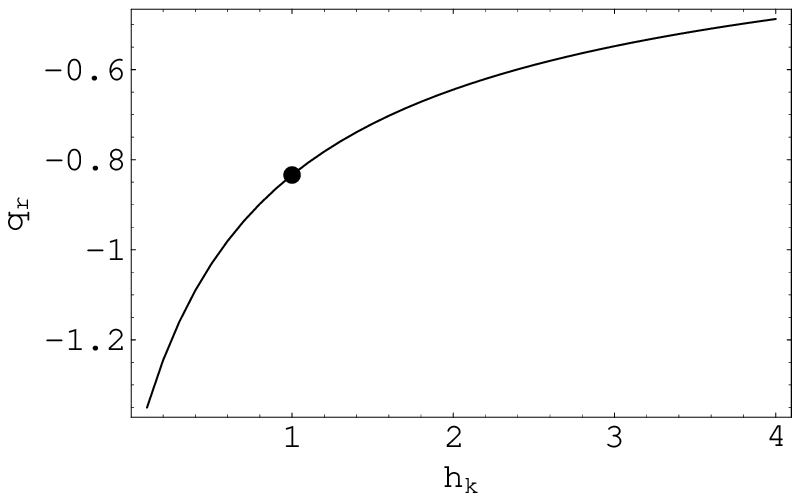}
  \caption{Illustration of the dependence of the deceleration parameter $q_r$
  on the parameters $(t,r_L,\rho_0,r_k,n_k,h_k)$,
  using the example in Table \ref{table:radial eg} as a reference
  (denoted by the large dot).}
  \label{fig:qrtopara}
\end{figure}

To study the dependence of the deceleration parameter $q_r$ on the
six parameters in Eq.\ (\ref{radial parameters}), we use the
example in Table \ref{table:radial eg} as a reference and tune one
of the six parameters at one time, while keeping the other five
unchanged (i.e., with the values in Table \ref{table:radial eg}).
The results are shown in Fig.\ \ref{fig:qrtopara}. The plot of
$q_r$ versus $n_k$ shows that we have larger acceleration for
larger $n_k$ that corresponds to larger inhomogeneity. This result
again supports the possibility of the inhomogeneity-induced
acceleration, in which, naively, inducing larger acceleration
requires larger inhomogeneity. In addition, we have acceleration
even for a moderate power $n_k$ (e.g., $n_k = 5$), that is, for
the purpose of generating the line acceleration we do not need
very large inhomogeneity.

\newpage

\section{Domain Acceleration induced by Inhomogeneities\label{dom}}
In this section we investigate the domain acceleration for a
spherical domain, $0<r<r_D$, with the volume
\begin{equation} \label{eq:ltb volume}
V_D %
= 4 \pi \int_0^{r_D} \frac{R^2 R,_{r}}{\sqrt{1+2E(r)}} \, dr %
= 4 \pi \int_0^{r_D} \frac{a^2 r^2 (a +
a,_{r}r)}{\sqrt{1-k(r)r^2}} \, dr \, .
\end{equation}
The length invoked for the domain acceleration is that in Eq.\
(\ref{domain size}): $L_D = V_D^{1/3}$, via which the deceleration
parameter $q_D$ of the domain is defined in Eq.\ (\ref{qD def}):
$q_D = -\ddot{L}_D L_D/\dot{L}_D^2 \,$. We note that in the case
where $k(r)=0=E(r)$ there is no domain acceleration (see Appendix
\ref{app:E=0} and Ref.\ \onlinecite{Paranjape:2006cd}).

%
%
For the domain acceleration Kai et al.\ \cite{Kai:2006ws} found
examples based on the LTB solution, where constant $\rho_0(r)$,
trivial $t_b(r)$ and the following $k(r)$ function are invoked.
\begin{equation}
 k(r)=
  \begin{cases}
   k_{0} & \text{for~~~} 0\leq r<r_{1}, \\
   \dfrac{k_{0}}{2r^{2}}
    \left\{
     \dfrac{(r^{2}-r_{2}^{2})^{2}}{r_{1}^{2}-r_{2}^{2}}
     +r_{1}^{2}+r_{2}^{2}
   \right\} & \text{for~~~} r_{1}\leq r<r_{2}, \rule{0pt}{20pt} \\
   \dfrac{k_{0}}{2r^{2}}
   \left(r_{1}^{2}+r_{2}^{2}\right)
   & \text{for~~~} r_{2}\leq r<r_{3}, \rule{0pt}{26pt} \\
   \dfrac{k_{0}}{2r^{2}}\left(r_{1}^{2}+r_{2}^{2}\right)
   \left\{
     \left(\dfrac{r^{2}-r_{3}^{2}}
      {r_{\rm b}^{2}-r_{3}^{2}}
     \right)^{2}-1
   \right\}^{2}
   & \text{for~~~} r_{3}\leq r<r_{\rm b}, \rule{0pt}{26pt} \\
   0 & \text{for~~~} r_{\rm b}\leq r, \rule{0pt}{26pt}
  \end{cases}
  \label{eq:kai}
\end{equation}
where $0<r_1<r_2<r_3<r_{\rm b}$ and $k_0$ is a constant.
In these examples \cite{Kai:2006ws}, the acceleration involves the
existence of a singularity around the origin.
In contrast, we find the domain acceleration examples without
singularity. The difference in the choice of the three functions
which specify the LTB solution is that in our examples non-trivial
$t_b(r)$ is invoked, as going to be presented in the following.
Our examples indicate that the existence of a singularity is not
necessary for generating the domain acceleration.
%

In search of examples of the domain acceleration without
singularity, we first follow the same procedures used in the
previous section for the line acceleration and choose the
functions, $k(r)$, $\rho_0(r)$, and $t_b(r)$, as those in Eqs.\
(\ref{k function})--(\ref{trivial tb function}). In this case we
surveyed the six-dimensional parameter space
$(t,r_L,\rho_0,r_k,n_k,h_k)$ and found no domain acceleration.

Contradicting our result, in Ref.\ \onlinecite{Nambu:2005zn} it
was claimed that the example of the domain acceleration was found
with constant $\rho_0$, trivial $t_b(r)$ and step-function-like
$k(r)$ [i.e., with infinitely large $n_k$ in Eq.\ (\ref{k
function})]. There is a mistake in the calculations of the volume
$V_D$ in Ref.\ \onlinecite{Nambu:2005zn}, where the authors
ignored the volume at the transition point $r=r_k$ that is
actually nonzero and should not be ignored (even though $r=r_k$
corresponds to a ``2D surface'' in the coordinate space). After
taking the volume at $r=r_k$ back into account, we found no domain
acceleration. (For more details, see Appendix
\ref{app:NT_example}.)

\begin{figure}[h]
  \center
  \includegraphics[width=0.65\linewidth,clip]{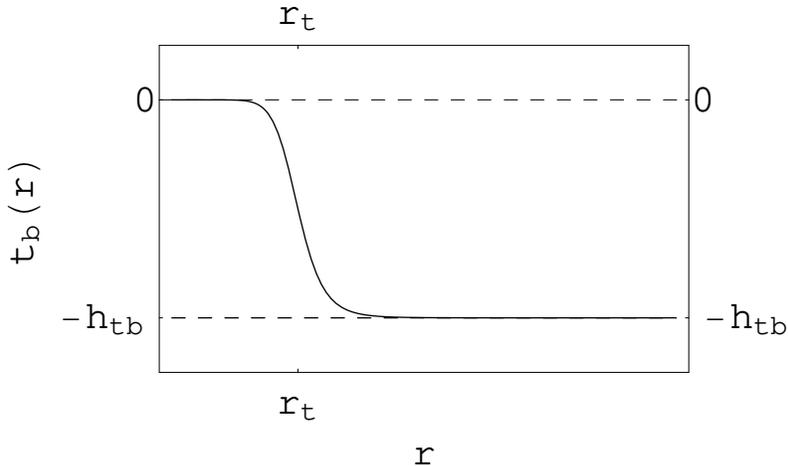}
  \caption{The plot of the function $t_b(r)$ invoked in the search for
  the domain acceleration.}
  \label{fig:tb(r)}
\end{figure}

Since no domain acceleration (without singularity) was found in
our search with trivial $t_b(r)$, we then consider a non-trivial
function for $t_b(r)\,$:
\begin{equation}
t_b(r) = - \frac{h_{tb}(r/r_t)^{n_t}}{1+(r/r_t)^{n_t}} \, ,
\label{nontrivial tb function}
\end{equation}
while invoking the same functions for $k(r)$ and $\rho_0(r)$ in
Eqs.\ (\ref{k function}) and (\ref{const rho0 function}):
\begin{eqnarray}
k(r) &=& -\frac{(h_k+1) (r/r_k)^{n_k}}{1+(r/r_k)^{n_k}}+1 \, , \\
\rho_0 (r) &=& \textrm{constant}  \, .
\end{eqnarray}
The behavior of the function $t_b(r)$ in Eq.\ (\ref{nontrivial tb
function}) is illustrated in Fig.\ \ref{fig:tb(r)}.

With the above choice of the three functions involved in the LTB
solution, we have nine free parameters to tune:
\begin{equation}
(t,r_D,\rho_0,r_k,n_k,h_k,r_t,n_t,h_{tb}) \, . %
\label{domain parameters}
\end{equation}
We surveyed this nine-dimensional parameter space
and did find examples of the domain acceleration eventually. In
Table \ref{table:domain eg}, we present three examples with
significantly different magnitude in acceleration (i.e., regarding
the value of $q_D$).

\tabcolsep=8pt
\begin{table}[h!]
\caption{Three examples of the domain acceleration.} \label{table:domain eg} %
\center
\begin{tabular}{|c|c|c|c|c|c|c|c|c|c||c|}\hline
&$t$ & $r_D$ & $\rho_0$ & $r_k$ & $n_k$ & $h_k$ & $r_t$ & $n_t$ & $h_{tb}$ & $q_D$ \\
\hline
1&$0.1$ & $1$ & $1$ & $0.6$ & $20$ & $10$ & $0.6$ & $20$ & $10$ & $-0.01$ \\
\hline
2&$0.1$ & $1.1$ & $10^5$ & $0.9$ & $40$ & $40$ & $0.9$ & $40$ & $10$ & $-1.08$ \\
\hline
3&$10^{-8}$ & $1$ & $10^{10}$ & $0.77$ & $100$ & $100$ & $0.92$ & $100$ & $50$ & $-6.35$ \\
\hline
\end{tabular}
\begin{tabular}{|c|c|c|c|c|}
\hline
& $L_D$ & $\dot{L}_D$ & $\ddot{L}_D$ & $q_D$\\
\hline
1&$16.2$&$1.62$&$0.00174$& $-0.01$\\
\hline
2& $94.0$ & $7.63$ & $0.694$ & $-1.08$ \\
\hline
3&$8720$&$117$&$10.0$& $-6.35$\\
\hline
\end{tabular}
\end{table}

\begin{figure}[tbph]
  \center
  \includegraphics[width=0.65\linewidth,clip]{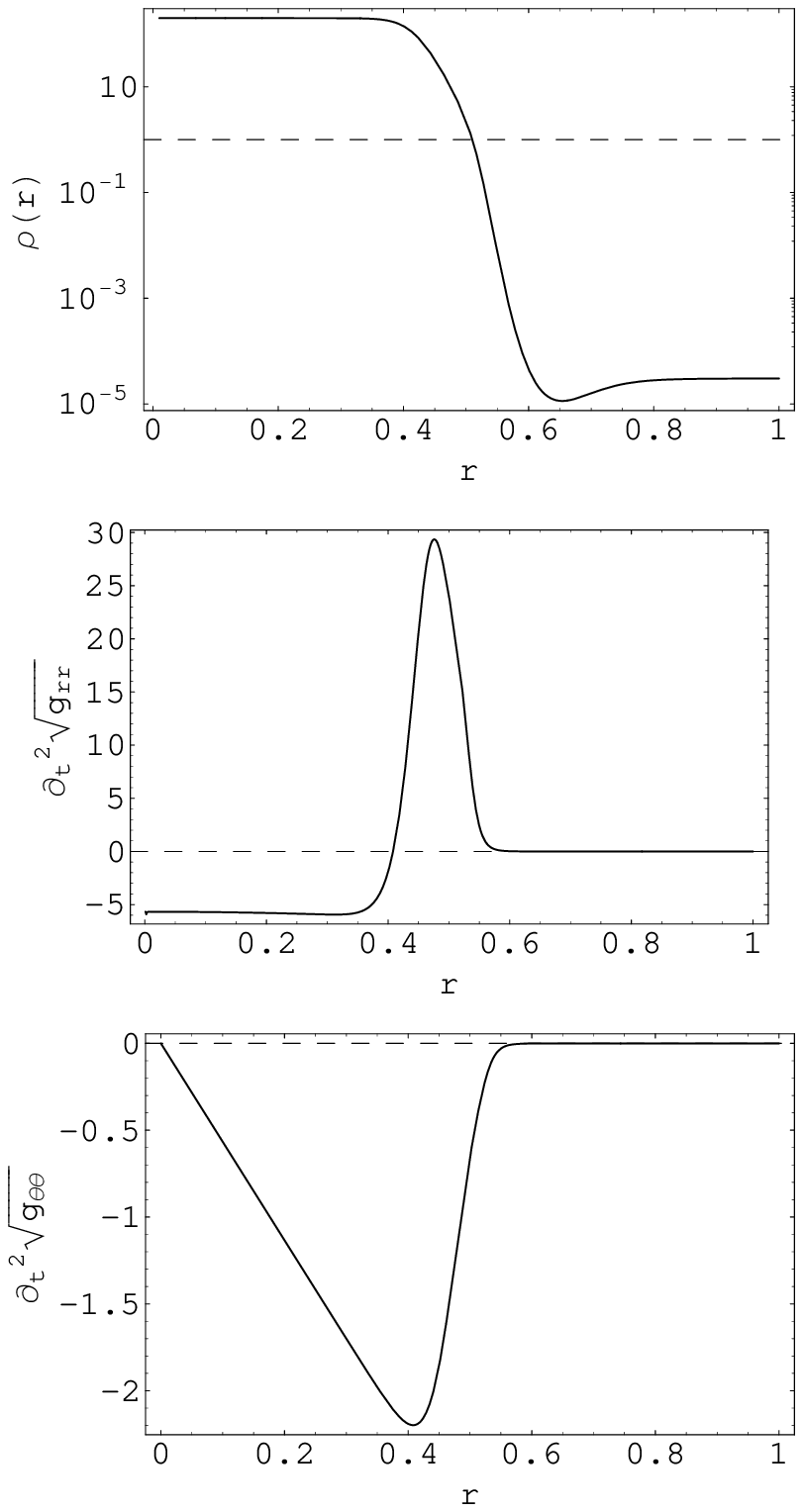}

  \caption{The plots of the physical energy density $\rho$ and the
  quantities, $\partial_t^2\sqrt{g_{rr}}$ and $\partial_t^2\sqrt{g_{\theta
  \theta}}$, which characterize the local acceleration/deceleration
  status in the radial and the angular direction, respectively,
  for the first example in Table \ref{table:domain eg}.
  Note that when $r=r_D=1$,
  $\partial_t^2\sqrt{g_{rr}}=-4.9\times10^{-5}$ and
  $\partial_t^2\sqrt{g_{\theta \theta}}=-1.6\times10^{-4}\,$.}
  \label{fig:qD1}
\end{figure}

\begin{figure}[tbph]
  \center
  \includegraphics[width=0.65\linewidth,clip]{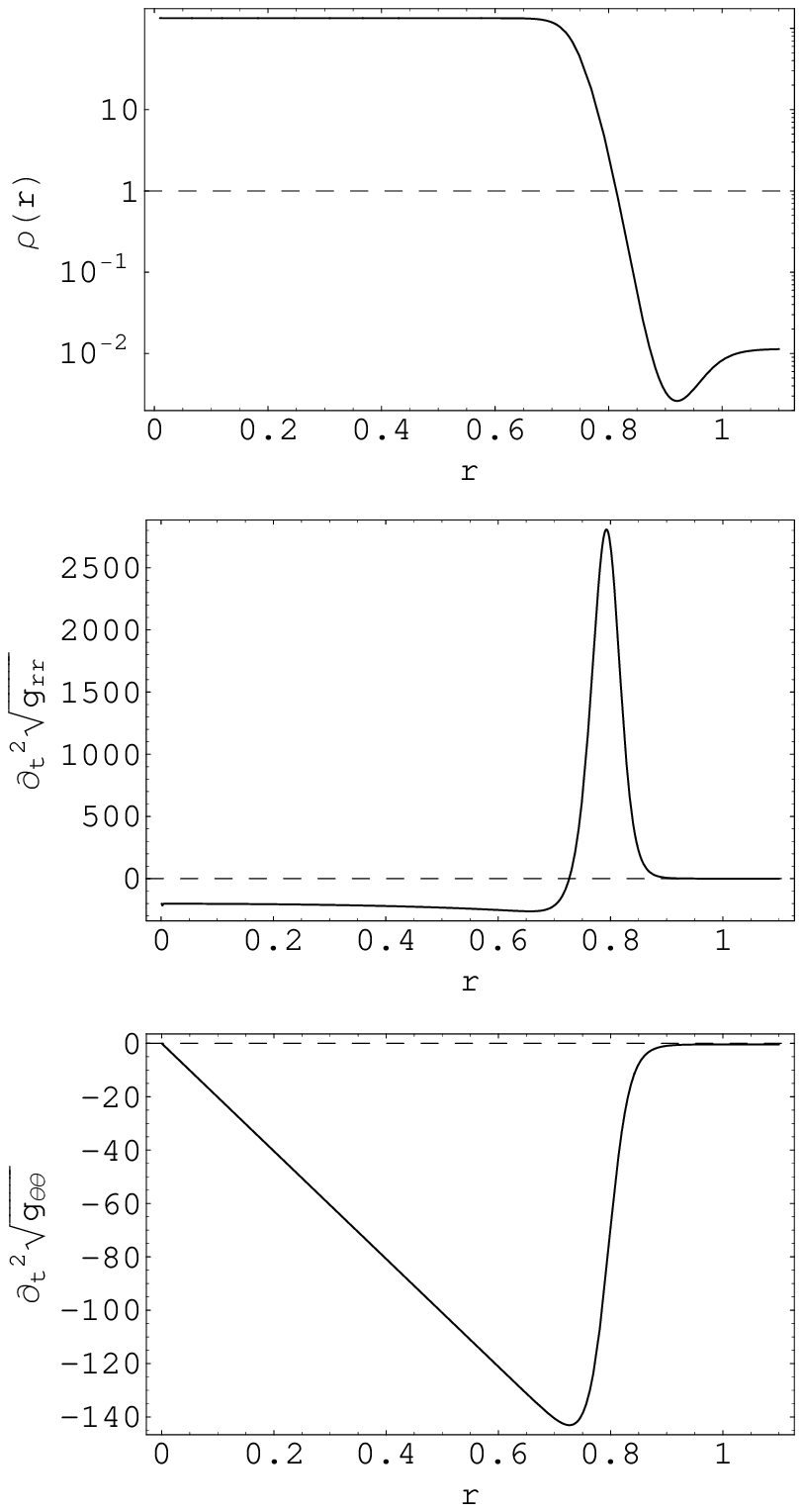}

  \caption{The plots of the physical energy density $\rho$ and the
  quantities, $\partial_t^2\sqrt{g_{rr}}$ and $\partial_t^2\sqrt{g_{\theta
  \theta}}$, which characterize the local acceleration/deceleration
  status in the radial and the angular direction, respectively,
  for the second example in Table \ref{table:domain eg}.
  Note that when $r=r_D=1.1$,
  $\partial_t^2\sqrt{g_{rr}}=-0.055$ and
  $\partial_t^2\sqrt{g_{\theta \theta}}=-0.43\,$.}
  \label{fig:qD2}
\end{figure}

\begin{figure}[tbph]
  \center
  \includegraphics[width=0.65\linewidth,clip]{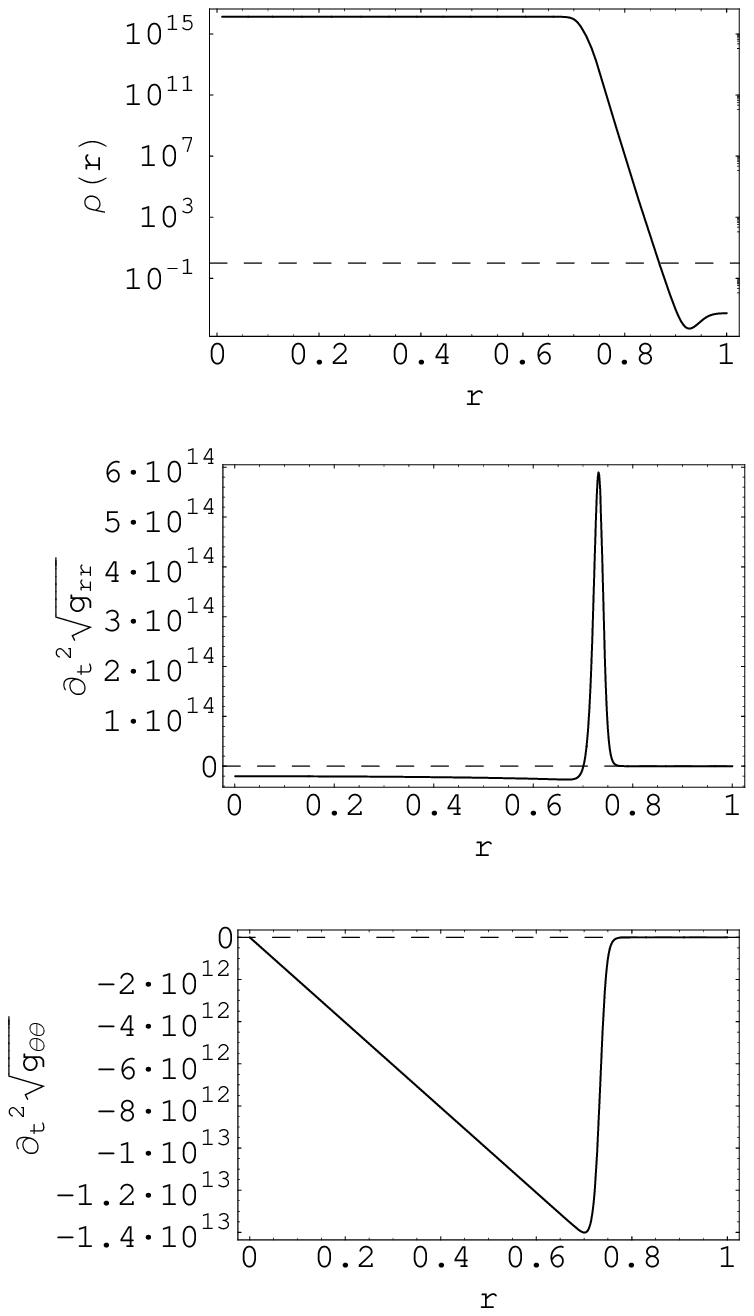}

  \caption{The plots of the physical energy density $\rho$ and the
  quantities, $\partial_t^2\sqrt{g_{rr}}$ and $\partial_t^2\sqrt{g_{\theta
  \theta}}$, which characterize the local acceleration/deceleration
  status in the radial and the angular direction, respectively,
  for the third example in Table \ref{table:domain eg}.
  Note that when $r=r_D=1$,
  $\partial_t^2\sqrt{g_{rr}}=-0.23$ and
  $\partial_t^2\sqrt{g_{\theta \theta}}=-2.36\,$.}
  \label{fig:qD3}
\end{figure}

The further details of these three examples are respectively
illustrated in Figs.\ \ref{fig:qD1}, \ref{fig:qD2}, and
\ref{fig:qD3}, which present the energy density distribution and
the acceleration/deceleration status of the radial and the angular
line elements. Similar to the example of the line acceleration in
the previous section, in all these three examples the spherically
symmetric dust fluid consists of two roughly homogeneous regions
and one inhomogeneous transition/junction region.\footnote{Note
that in these three acceleration examples the energy density
distribution is flat around the origin $r=0$ and therefore there
is no singularity and, moreover, no cusp behavior (or weak
singularity).} Regarding the line elements, we get acceleration in
the radial direction in the inhomogeneous region. How the
evolution of the line elements affect the evolution of $L_D$ is
not clear. Naively, it is a reasonable possibility that the domain
acceleration stems from the acceleration of the radial line
elements in the inhomogeneous region, that is, the existence of
the accelerating line elements might be a necessary condition for
the domain acceleration. Is it possible to have the domain
acceleration without the acceleration of line elements? So far we
do not have a no-go theorem prohibiting this possibility. However,
we found no such example through our survey of the
nine-dimensional parameter space. This might be an indication of
the correlation between the domain acceleration and the
acceleration of line elements.
%


For a demonstration of the scales of the size and other quantities
of this system, in Table IV 
we fix the only one unspecified unit by using the length unit:
$0.1$Mpc, $1$Mpc, $10$Mpc and $100$Mpc, and present the values of
several dimensionful quantities (in addition to the dimensionless
$q_D$), respectively. %
For each example in Table \ref{table:domain eg}, when the size of
the system increases by one order of magnitude, the time increases
by one order and the energy density decreases by two orders of magnitude. %
From this table one can see that the time $t$ and the energy
density of the outer region cannot be simultaneously consistent
with the situation of the present universe (i.e., $t_0 \sim
10^{10}\,$years and $\rho_0 \sim 10^{-29}\,$g/cm$^{-3}$), and
therefore these three examples by themselves cannot describe the
present universe. The search (based on the LTB solution) of the
domain acceleration examples which are consistent with
observational results is important and worthy of further
investigations.

\begin{table}[h!] 
\caption{Corresponding to different length units, the values of
several dimensionful quantities for the acceleration examples in
Table \ref{table:domain eg} are presented.} %
\center
\begin{tabular}{|c|c|c|c|c|c|c|}\hline
& $q_D$ & Length unit & $t$ (year) & $L_r$ (Mpc)
& $\rho(r=0)$ (g/cm$^3$) & $\rho(r=r_D)$ (g/cm$^3$) 
\\ \hline 1 &$-0.01$& 0.1 Mpc &3.26E4 &1.62  &3.15E-30 &4.81E-37 
\\ \hline   &       & 1 Mpc   &3.26E5 &16.2  &3.15E-32 &4.81E-39 
\\ \hline   &       & 10 Mpc  &3.26E6 &162  &3.15E-34 &4.81E-41 
\\ \hline   &       & 100 Mpc &3.26E7 &1620 &3.15E-36 &4.81E-43 
\\ \hline 2 &$-1.08$& 0.1 Mpc &3.26E4   &9.4  &2.11E-30 &1.79E-34 
\\ \hline   &       & 1 Mpc   &3.26E5  &94  &2.11E-32 &1.79E-36 
\\ \hline   &       & 10 Mpc  &3.26E6  &940 & 2.11E-34 &1.79E-38 
\\ \hline   &       & 100 Mpc &3.26E7  &9400  & 2.11E-36 &1.79E-40 
\\ \hline 3 &$-6.35$&0.1 Mpc &3.26E-3  &872  &2.11E-16 &8.33E-36 
\\ \hline   &       &1 Mpc   &3.26E-2   &8720  & 2.11E-18 &8.33E-38 
\\ \hline   &       &10 Mpc  &3.26E-1  &87200  &2.11E-20 &8.33E-40 
\\ \hline   &       &100 Mpc &3.26  &872000  &2.11E-22 &8.33E-42 
\\ \hline
\end{tabular}
\end{table}

\begin{figure}[tbph]
  \centering
  \includegraphics[width=0.4\linewidth,clip]{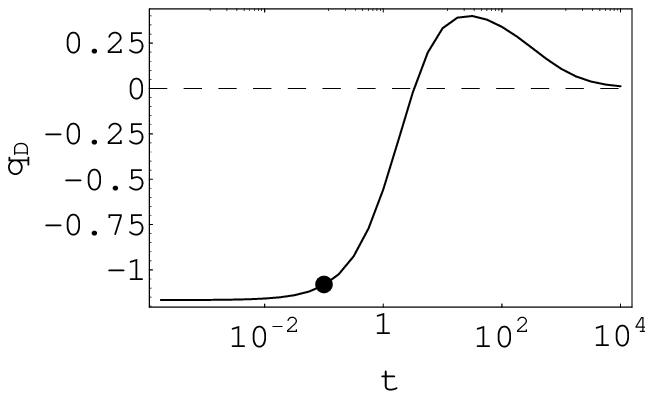}
  \includegraphics[width=0.4\linewidth,clip]{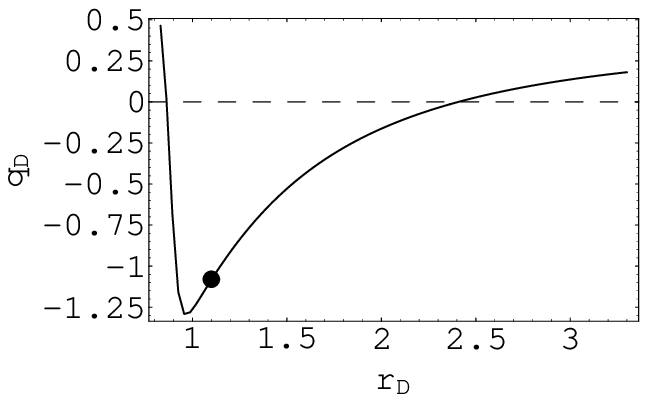}
  \includegraphics[width=0.4\linewidth,clip]{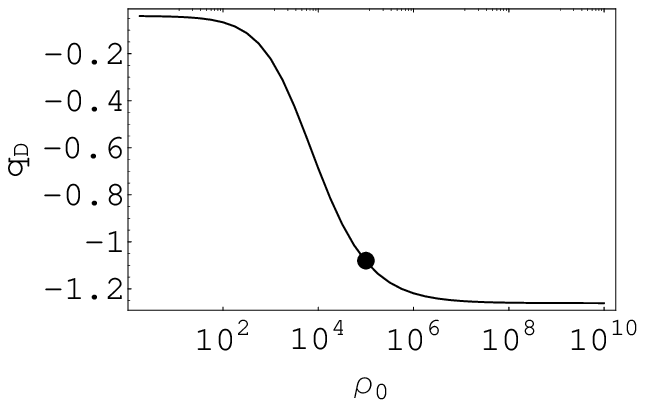}
  \includegraphics[width=0.4\linewidth,clip]{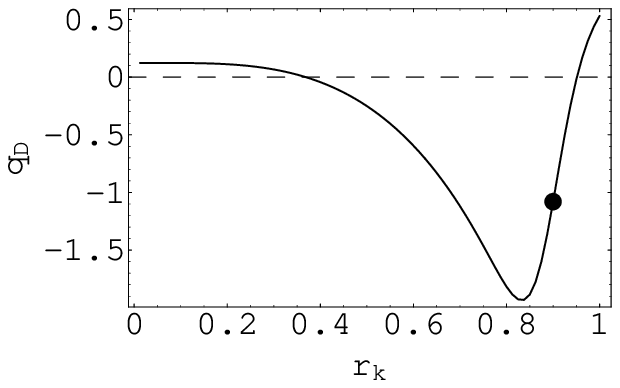}
  \includegraphics[width=0.4\linewidth,clip]{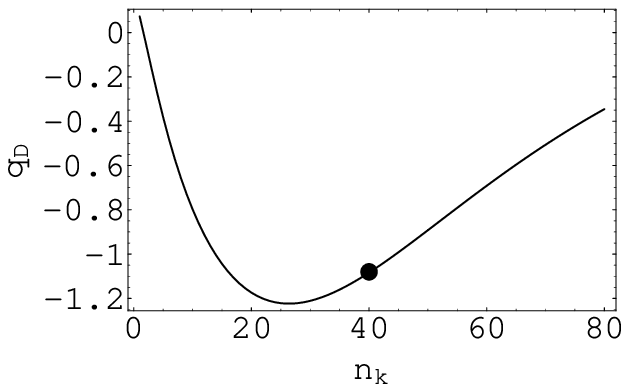}
  \includegraphics[width=0.4\linewidth,clip]{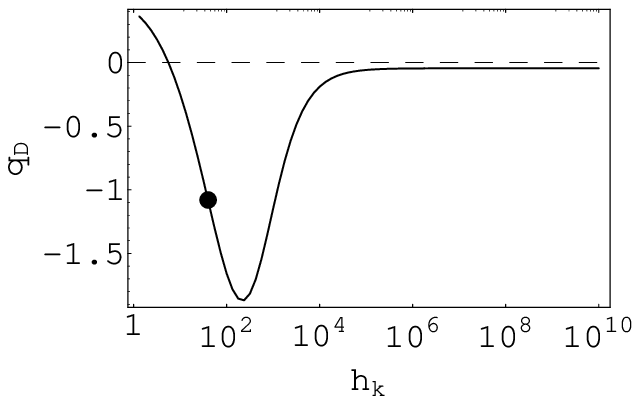}
  \includegraphics[width=0.4\linewidth,clip]{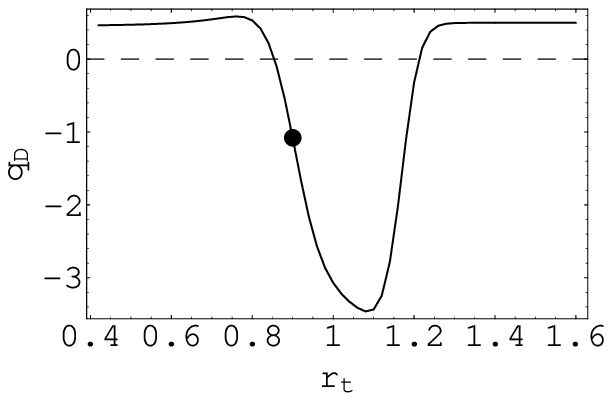}
  \includegraphics[width=0.4\linewidth,clip]{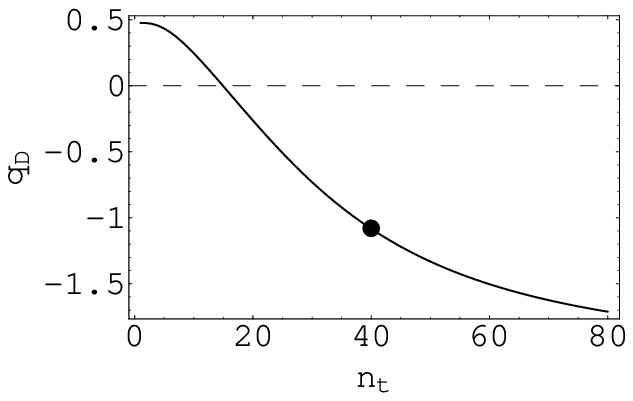}
  \includegraphics[width=0.4\linewidth,clip]{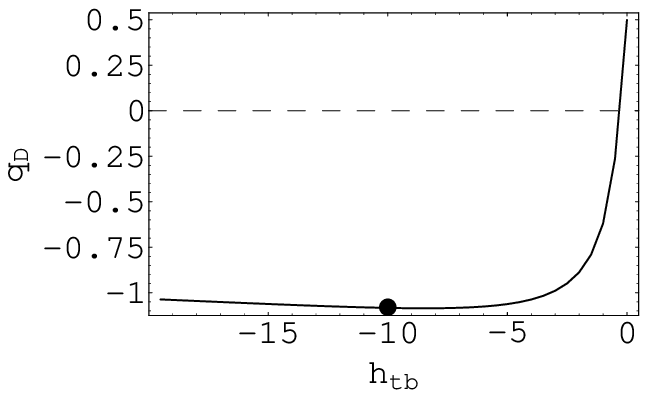}

  \caption{Illustration of the dependence of the deceleration parameter $q_D$
  on the parameters $(t,r_D,\rho_0,r_k,n_k,h_k,r_t,n_t,h_{tb})$,
  using the second example in Table \ref{table:domain eg} as a
  reference (denoted by the large dot).}
  \label{fig:qdtopara}
\end{figure}

To study the dependence of the deceleration parameter $q_D$ on the
nine parameters in Eq.\ (\ref{domain parameters}), we use the
second example in Table \ref{table:domain eg} as a reference and
tune one of these nine parameters at one time, while keeping the
other eight unchanged (i.e., with the values in the second example
in Table \ref{table:domain eg}). The results are shown in Fig.\
\ref{fig:qdtopara}. These plots show that every parameter has
significant influence on $q_D$ (at least within some range of the
value of the parameter). We note that in the case of the
inhomogeneous function $t_b(r)$ in Eq.\ (\ref{nontrivial tb
function}) we have acceleration or deceleration in different
situations with different values of parameters, and, moreover, the
examples of the domain acceleration are not rare. Conversely, in
the case of trivial $t_b(r)$ we found no domain acceleration
without singularity. Accordingly, in the LTB solution the function
$t_b(r)$ plays a key role in generating the domain acceleration.


\section{Summary, Discussions and Outlook} \label{con}

Against the common consensus [in Eq.\ (\ref{consensus})] that
normal matter always slows down the expansion of the universe, we
have found and demonstrated examples 
of the line acceleration and the domain acceleration for a
spherically symmetric dust fluid described by the LTB solution
without singularity.
This discovery contradicts the common intuition about the
interplay of gravity and the cosmic evolution. Furthermore, these
examples have shown the strong correlation between acceleration
and inhomogeneity. These results strongly support the suggestion
that inhomogeneity can induce acceleration.

In these acceleration examples the spherically symmetric dust
fluid consists of three regions: two (roughly) homogeneous regions
--- the inner over-density region and the outer under-density
region --- and one transition/junction region, where the
inhomogeneity locates, of two homogeneous regions. Note that in
these examples the energy density distribution is flat around the
origin $r=0$ and therefore there is no singularity and, moreover,
no cusp behavior (or weak singularity).

For further understanding the acceleration, the quantity
$\partial_t^2 \sqrt{g_{rr}}\,$, which characterizes the
acceleration/deceleration status of an infinitesimal radial line
element, is one of the good quantities to study. Naively, the
regions with positive/negative $\partial_t^2 \sqrt{g_{rr}}$ make
positive/negative contribution to acceleration.
For the quantity $\partial_t^2 \sqrt{g_{rr}}$ to be positive we
find a necessary and sufficient condition: $(a^2/r)_{,r}>0$. This
condition tells us that a positive contribution to acceleration is
made in the place where $a(t,r)$ increases sufficiently fast with
the radial coordinate $r$.
In every acceleration example we found, there exists a region with
positive $\partial_t^2 \sqrt{g_{rr}}$ that coincides with the
inhomogeneous transition/junction region quite well. This result
reveals the strong correlation between acceleration and
inhomogeneity, thereby giving a strong support to the suggestion
that inhomogeneity can induce accelerating expansion.


It is not clear how the line acceleration and domain acceleration
are related to the apparent acceleration. A similar doubt
(regarding the domain acceleration) was also raised by Kai et al.\
in Ref.\ \onlinecite{Kai:2006ws}. The relation between the
theoretical accelerations --- line acceleration and domain
acceleration --- and the apparent acceleration could be
model-dependent. For our universe with complicated energy
distribution the relation may be far from simple. %
As a reminder, we note that the conclusion about the existence of
the present cosmic acceleration indicated by observations is based
on the homogeneous and isotropic FLRW cosmology. If one invokes
inhomogeneous cosmology, the current observational results may not
indicate the existence of the cosmic acceleration. %
As shown in Appendix \ref{app:E=0} and Ref.\
\onlinecite{Paranjape:2006cd}, in the cases where $E(r)=0$ there
is no acceleration (for both the line and the domain
acceleration). In contrast, it has been shown that it is possible
to fit supernova data in the LTB model with trivial $E(r)$
\cite{Celerier:1999hp}. This gives an example of having apparent
acceleration while having no line acceleration and no domain
acceleration. %
One may treat a complicated universe as a large domain consisting
of many different sub-domains, and from a statistical perspective
it might be reasonable that the size $L_D$ of the large domain
corresponds to the scale factor in the FRW metric. In many works
the results from the analysis involving the quantity $L_D$ are
compared with observations in this spirit
\cite{Buchert:1999pq,Rasanen:2006zw,Rasanen:2006kp,Wiltshire:2007fg,Li:2007ny,Rasanen:2008it}.
Nevertheless, whether this is a good approximation is not clear
yet.


One may wonder whether the acceleration in our examples is real or
fake, i.e., whether it truly corresponds to the space expansion.
One example leading to this concern is to consider an accelerating
system consisting of two decelerating regions. Since each region
is decelerating one might expect that the domain acceleration in
this case does not correspond to physically observable attributes
(e.g.\ the luminosity distance-redshift relation) of an
accelerating FLRW model \cite{Ishibashi:2005sj}. Nevertheless, on
the contrary, it has been shown that for a system consisting of
decelerating regions it is possible to fit the supernova data or
to generate apparent acceleration
\cite{Celerier:1999hp,Kolb:2005ze}.

In addition, as already pointed out in Introduction, it needs much
caution to connect separate regions or to put them together. Two
particular important issues (that may be closely related to each
other) are: \\
(1) There may be singularity in the junction between separate
regions. \\
(2) The effect from the junction between separate regions may be
significant and should be taken into account seriously. \\
We have already had a lesson from the work by Nambu and Tanimoto
in Ref.\ \onlinecite{Nambu:2005zn}, where they basically
considered a system (of spherical symmetry) consisting of an inner
and an outer decelerating FLRW region and, as a result, they found
and presented examples of the domain acceleration. In this system
there is singularity in the junction between these two regions,
and in their treatment the contribution from the junction is
ignored. As pointed out in the previous section (with details in
Appendix \ref{app:NT_example}), it is inappropriate to ignore the
junction. Actually the junction makes significant contribution to
the volume of the system, although it looks like a ``2D'' surface
in the coordinate space. After taking care of the singularity and
taking into account the volume of the junction, we find that there
exists no domain acceleration in the cases studied in Ref.\
\onlinecite{Nambu:2005zn}.

Another concern is about the definition of acceleration, as
discussed in Sec.\ \ref{def}.
The two definitions employed in the present paper --- line
acceleration and domain acceleration, on which our acceleration
examples are based --- follow the scenario proposed in Sec.\
\ref{def} for a system of freely moving particles interacting with
each other only through gravity. For the purpose of truly
representing the space evolution status while avoiding the
confusion from particle motion and fake frame acceleration, the
length quantities invoked in this scenario to define acceleration
are (1) the distance between two freely moving particles for the
line acceleration and (2) the size of a spatial region with
constant number of particles therein for the domain acceleration.
Regarding the frame/gauge choice, we have emphasized the advantage
of the synchronous gauge in which the above requirements for the
length involved in the acceleration definition can be easily met.
In addition, in this gauge, there exists a universal cosmic time
that is the proper time of comoving observers, and for a region
with its boundary fixed in the coordinate space the volume
expansion rate coincides with the expansion rate of the physical
proper volume.


The counter-intuitive acceleration examples we found raise two
issues worthy of further investigations: %
\vspace{0.1em}

\noindent %
\hspace*{0em} $\bullet$ \textit{How to understand these counter-intuitive examples?}\\ %
\hspace*{0em} $\bullet$ \textit{Can inhomogeneities explain ``cosmic acceleration''?} %
\vspace{0.3em}

Regarding the first issue, the common intuition about Eq.\
(\ref{consensus}) may actually stem from Newtonian gravity that is
a perturbed version of general relativity in the Newtonian limit
with the Minkowski space-time as the background. Accordingly, this
intuition may be valid only for considering the particle motion
relative to a background space-time in a perturbative framework,
but may be invalid for considering the evolution of space-time
that is described by general relativity. For further understanding
the cosmic evolution and other topics involving the
general-relativity effects as the dominant effects, it will be
helpful if one can build the intuition about general relativity,
i.e., with which one can make a proper guess at the behavior of
the space-time geometry for an arbitrarily given energy
distribution.



Regarding the second issue, the most important is whether the
inhomogeneities of our universe can explain the supernova data. If
yes, the next step is to see, according to observational data, how
the universe evolves, in particular, whether the cosmic
acceleration exists or not, in this cosmological model taking the
real inhomogeneities into account. So far the examples we found
may be far away from the real situation of our universe. How to
benefit from these mathematical examples in order to understand
the present cosmic acceleration through the inhomogeneities of our
universe is an important issue currently under our investigations.
No matter whether the comic acceleration can eventually be
explained simply by inhomogeneity, these examples have shown that
inhomogeneity may affect the cosmic evolution in a manner far
beyond the usual naive intuition about the interplay of gravity
and the space-time geometry. Thus, the effects of inhomogeneity on
the cosmic evolution should be restudied carefully. These examples
open a new perspective on the understanding of the evolution of
our universe.


\begin{acknowledgments}
We thank P.-M.\ Ho for useful discussions. This work was supported
by the National Science Council, Taiwan, R.O.C.\ (NSC
94-2752-M-002-007-PAE, NSC 94-2112-M-002-029, NSC
94-2811-M-002-048, NSC 94-2112-M-002-036, and NSC
93-2112-M-002-009). Gu also thanks the National Center for
Theoretical Sciences (funded by the National Science Council),
Taiwan, R.O.C.\ for the support.
\end{acknowledgments}


\appendix

\section{Restrictions on the LTB Solution}\label{app:ltb_constraint}

In this appendix, we will give several reasonable restrictions on
the LTB solution and study the properties of the LTB solution
under these restrictions. We consider three conditions: (1) no
hole, (2) no singularity, and (3) non-negative and finite energy
density. For no hole at the center, the area of the spherical
surface at $r = r_0$ should go to zero when $r_0$ goes to zero.
Accordingly, Condition (1) requires
\begin{equation}\label{ltb_constraint1}
R(t,r=0)=0 \, .
\end{equation}
Condition (2) requires the smooth behavior of the functions
involved in the LTB solution. In particular, we consider
\begin{equation}\label{ltb_constraint2}
R,_r (t,r)  \neq 0 \, .
\end{equation}
Condition (3) requires
\begin{equation}\label{ltb_constraint3}
0 \leq \rho (t,r) < \infty \, .
\end{equation}
Note that the necessary and sufficient conditions of no shell
crossing in a period of time are presented by Hellaby and Lake in
Ref.\ \onlinecite{Hellaby:1985zz}.
The conditions discussed here are only necessary conditions.

Because $R,_r$ is nonvanishing and continuous, $R,_r$ cannot
change its sign, i.e., $R,_r$ is either always positive or always
negative for all $r$. Because $R(t,r=0)=0$, the sign of $R(t,r>0)$
should be the same as that of $R,_r$, i.e.,
\begin{equation}\label{ltb_constraint4}
R R,_r > 0 \quad \mbox{ for } \; r > 0 \, .
\end{equation}
This property will be used in Appendix\ \ref{app:NT_example}
regarding the possibility of generating the domain acceleration
with trivial $t_b(r)$.

Considering Eq.\ (\ref{eq2}),
\begin{equation}
\left({\frac{\dot{R}}{R}}\right)^2 = \frac{2
E(r)}{R^2}+\frac{2M(r)}{R^3} \, , \nonumber
\end{equation}
for every term therein to be finite, we obtain
\begin{eqnarray}
E(r=0) &=& 0 \, , \label{ltb_constraint5} \\
M(r=0) &=& 0 \, , \label{ltb_constraint6}
\end{eqnarray}
because $R(t,r=0)=0$.
From Eq.\ (\ref{eq3}),
\begin{equation}
\rho(t,r) = \frac{2M'(r)}{R^2 R,_{r}} \, , \nonumber
\end{equation}
Eq.\ (\ref{ltb_constraint3}) requires
\begin{equation}\label{ltb_constraint7}
\frac{M'(r)}{R,_r(t,r)} \geq 0 \, .
\end{equation}
Because $R,_r$ cannot change its sign, $M'(r)$ cannot change its
sign either according to the above relation. Using the same
reasoning for obtaining Eq.\ (\ref{ltb_constraint4}), from Eq.\
(\ref{ltb_constraint6}) we have
\begin{equation}\label{ltb_constraint8}
M(r)M'(r) \geq 0 \, .
\end{equation}

To sum up, so far we have obtained the following basic
restrictions on the LTB solution: $R(t,r=0)=E(r=0)=M(r=0)=0$; $R$,
$R,_r$, $M(r)$ and $M'(r)$ have the same sign when nonvanishing.

Examining Eqs.\ (\ref{eq2}) and (\ref{eq3}), we can see the
feature that for a solution where $R(t,r) \leq 0$ there is always
another solution $R_1(t,r) \geq 0$, obeying
\begin{equation}\label{ltb_constraint9}
R_{1}(t,r)=-R(t,r), \; M_{1}(r)=-M(r), \; E_{1}(r)=E(r) \, .
\end{equation}
Thus, without losing generality, we can set
\begin{equation}\label{ltb_constraint10}
R(t,r) \geq 0, \; 
R,_r > 0, \; M(r) \geq 0, \; M'(r) \geq 0 \, .
\end{equation}
With this choice, regarding the dependence on the radial
coordinate $r$, $R$ is a monotonically increasing function and $M$
is a non-decreasing function, while both vanish at the origin
$r=0$.

In the following we will study $\ddot{R}(t,r)$ and show that
$\ddot{R}(t,r) \leq 0$ and $\ddot{R}(t,r=0)=0$ corresponding to
the choice in Eq.\ (\ref{ltb_constraint10}). The time derivative
of Eq.\ (\ref{eq2}) multiplied by $R^2$ gives a simple formula for
$\ddot{R}$:
\begin{equation}\label{ltb_constraint11}
\ddot{R}=-\frac{M(r)}{R^2} < 0 \, ,
\end{equation}
corresponding to the choice in Eq.\ (\ref{ltb_constraint10}).
According to Eqs.\ (\ref{ltb_constraint1}) and
(\ref{ltb_constraint6}) and applying L'Hospital's rule, we have
\begin{equation}\label{ltb_constraint12}
\ddot{R}(t,r=0) = - \left. \frac{M}{\,R^2} \right|_{r=0} = -
\left. \frac{M'}{2 R R,_r} \right|_{r=0} = \left. - \frac{1}{4} R
\rho \right|_{r=0} \, ,
\end{equation}
where Eq.\ (\ref{eq3}) has been applied in the last equality. As a
result, for a finite physical energy density $\rho$,
\begin{equation}\label{ltb_constraint16}
\ddot{R}(t,r=0)=0 \, ,
\end{equation}
because $R(t,r=0)=0$.

\section{No Acceleration with $E(r)=0=k(r)$}\label{app:E=0}

In this appendix we consider the trivial case, $E(r)=0=k(r)$, and
prove that there is no line acceleration and no domain
acceleration in this case. (For the case of domain acceleration,
see also Ref.\ \onlinecite{Paranjape:2006cd}.)

\subsection*{Line Acceleration}
In the case with trivial $E(r)$ or $k(r)$, Eqs.\ (\ref{eq:lr}) and
(\ref{eq:grr}) give
\begin{equation}
L_r(t) \equiv \int_0^{r_L} \sqrt{g_{rr}}dr = \int_0^{r_L}R,_{r}dr
= R(t,r_L) \, ,
\end{equation}
where Eq.\ (\ref{ltb_constraint1}), $R(t,r=0)=0$, has been used,
and therefore
\begin{equation} \label{eq:qr:E=0}
q_r \equiv - \frac{\ddot{L}_r L_r}{\dot{L}_r^2} = - \left.
\frac{\ddot{R}R}{\dot{R}^2} \right|_{r=r_L} \, .
\end{equation}
From Eq.\ (\ref{eq2}), we have
\begin{equation} \label{eq:dotR:E=0}
\dot{R}^2 = \frac{2M(r)}{R} \, ,
\end{equation}
and
\begin{equation} \label{eq:ddotR:E=0}
\ddot{R} = - \frac{M(r)}{R^2} \, .
\end{equation}
Substituting these two equations into Eq.\ (\ref{eq:qr:E=0}) gives
a simple result:
\begin{equation}
q_r = \frac{1}{2} \, .
\end{equation}
Thus, in the case with trivial $E(r)$ or $k(r)$, no matter how we
tune the other two functions $\rho_0(r)$ and $t_b(r)$, the
deceleration parameter corresponding to the proper distance
between the origin and any other point in space is always $1/2$, a
positive constant denoting a deceleration as large as that in a
homogeneous dust-dominated or pressureless-matter-dominated universe. %

\subsection*{Domain Acceleration}

When $E(r)=0=k(r)$, Eq.\ (\ref{eq:ltb volume}) gives
\begin{equation}
V_D = 4 \pi \int_0^{r_D} R^2 R,_{r} dr = \frac{4\pi}{3} R^3(t,r_D)
\, ,
\end{equation}
\begin{equation}
L_D \equiv V_D^{1/3} = \left( \frac{4\pi}{3} \right)^{1/3}
R(t,r_D) \, ,
\end{equation}
and therefore
\begin{equation}
q_D \equiv - \frac{\ddot{L}_D L_D}{\dot{L}_D^2} = - \left.
\frac{\ddot{R}R}{\dot{R}^2} \right|_{r=r_D} = \frac{1}{2} \, ,
\end{equation}
where Eqs.\ (\ref{eq:dotR:E=0}) and (\ref{eq:ddotR:E=0}) have been
applied in the last equality.

Thus, the deceleration parameters corresponding to the domain
acceleration and the line acceleration are both $1/2$, a positive
constant (denoting deceleration) independent of the other two
functions $\rho_0(r)$ and $t_b(r)$. This indicates that a
non-trivial $E(r)$ or $k(r)$ function must play an essential role
in the inhomogeneity-induced accelerating expansion based on the
LTB solution.

\section{Domain Acceleration with Trivial $t_b(r)$?}\label{app:NT_example}

In \cite{Nambu:2005zn} Nambu and Tanimoto claimed that the example
of the domain acceleration was found. However, we find that it is
an improper example. In this appendix we will study the example in
\cite{Nambu:2005zn} and discuss its problems.

\subsection{Nambu and Tanimoto's Example}

In \cite{Nambu:2005zn} the choice of the three arbitrary functions
involved in the LTB solution are specified as follows:
\begin{eqnarray}
\rho_0(r) &=& \rho_0 = \textrm{constant} \, , \\
t_b(r) &=& 0 \, , \\
\label{eq:NT-k(r)}  k(r) &=&
\frac{1}{L^2}\left[2\theta(r-r_k)-1\right], \quad 0 \leq r \leq L,
\quad 0 \leq r_k \le L ,
\end{eqnarray}
where $\theta (r)$ is a step function. In the following study we
will choose $L=1$ without losing generality, that is,
\begin{equation}
\label{eq:NT-k(r):L=1}  k(r)=2\theta(r-r_k)-1, \quad 0 \leq r \leq
1, \quad 0 \leq r_k \leq 1 .
\end{equation}

With the above choice the spherically symmetric dust fluid
described by the LTB solution seems to consist of two regions
described respectively by two different Robertson-Walker (RW)
metrics: the inner region ($0 \leq r < r_k$) described by a
(spatially) open RW metric with the scale factor $a_1(t)$ and the
outer region ($r_k < r \leq 1$) described by a closed RW metric
with the scale factor $a_2(t)$. The scale factors $a_1(t)$ and
$a_2(t)$ obey the following Friedmann equations, respectively:
\begin{eqnarray}
  \left(\frac{\dot{a}_1}{a_1}\right)^2 &=&
  \frac{1}{a_1^2} + \frac{\rho_0}{3a_1^3} \, , \label{eq:NT-a1 eq}  \\
  \left(\frac{\dot{a}_2}{a_2}\right)^2 &=&
  -\frac{1}{a_2^2} + \frac{\rho_0}{3a_2^3} \, . \label{eq:NT-a2 eq}
\end{eqnarray}
The initial condition
\begin{equation}
a_1(t=0)=a_2(t=0)=0 \label{eq:NT-initial condition}
\end{equation}
are used to solve the above equations and to obtain the time
evolution of the scale factors ($a_1$ and $a_2$) and accordingly
the time evolution of the domain $\{ 0<r<L=1 \}$.

To obtain the deceleration parameter $q_D$ of this domain, the
following formulae for calculating the volume of the domain are
used:
\begin{eqnarray}
  V_D(t)&=&4 \pi \int_0^1
  \frac{a^2 r^2 \left( a+a,_{r}r \right) dr}{\sqrt{1-k(r)r^2}}
  \nonumber \\
  &=&4 \pi \left[ c_1a_1^3(t)+c_2a_2^3(t) \right] \label{eq:NT-volume} ,
\end{eqnarray}
where $c_1$ and $c_2$ are constants defined as follows:
\begin{eqnarray} \label{eq:NT-c1&c2}
  c_1=\int_0^{r_k}\frac{x^2dx}{\sqrt{1+x^2}} \, , \quad
  c_2=\int_{r_k}^1\frac{x^2dx}{\sqrt{1-x^2}} \, .
\end{eqnarray}
This volume calculation (i.e., the total volume being equal to the
sum of the volumes of the inner region and the outer region) looks
reasonable, but actually is incorrect, as to be discussed later.

\begin{figure}[h]
  \centering
  \includegraphics[width=0.5\linewidth,clip]{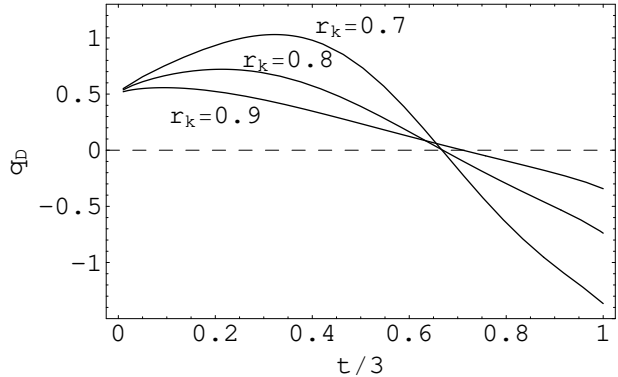} %
  \caption{Evolution of the deceleration parameter $q_D$ of a spherically symmetric
  domain, which consists of an inner open RW region and an outer closed RW region,
  for $\rho_0=3$ and $r_k=0.7,0.8,0.9$.}
  \label{fig:ntw}
\end{figure}

For demonstration, in the following we consider the special case
with $\rho_0 = 3$. The result about the time evolution of $q_D$,
obtained by using Eqs.\ (\ref{eq:NT-a1 eq})--(\ref{eq:NT-c1&c2}),
is shown in Fig.\ \ref{fig:ntw}. This figure illustrates the
existence of the domain acceleration at later times for several
different values of $r_k$.

\begin{figure}[h]
  \centering
  \includegraphics[width=0.5\linewidth,clip]{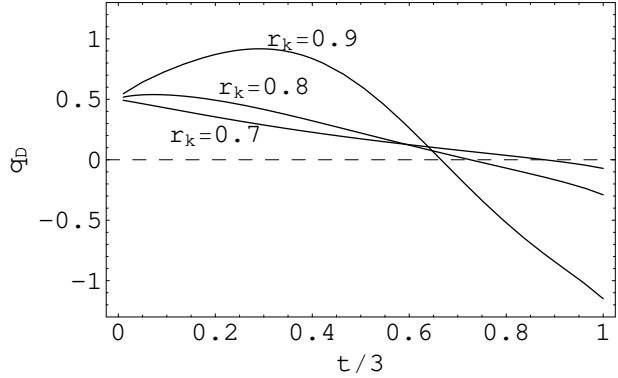}%
  \caption{Evolution of the deceleration parameter $q_D$ of a spherically symmetric
  domain, which consists of an inner closed RW region and an outer open RW region,
  for $\rho_0=3$ and $r_k=0.7,0.8,0.9$.}
  \label{fig:ntc}
\end{figure}

Note that with the initial condition in Eq.\ (\ref{eq:NT-initial
condition}) we have $a_1(t) > a_2(t)$ as $t>0$. This contradicts
the restriction in Eq.\ (\ref{ltb_constraint4}) around $r=r_k$
(i.e., around the junction of two regions), and accordingly
indicates the existence of a singularity that stems from the
increasing behavior of $k(r)$ around $r_k$ [but is irrelevant to
the singular behavior of $k(r)$ around $r_k$]. This singularity
can be avoided by exchanging the inner region and the outer
region, making the function $k(r)$ decreasing instead of
increasing, which is what we did about the choice of $k(r)$ in our
search for acceleration examples.

Accordingly, with this change, $k(r)$ becomes:
\begin{equation} \label{eq:NT-k(r):L=1:exchange}
  k(r) = -2\theta(r-r_k)+1,\quad 0 \leq r \leq
  1 , \quad 0 \leq r_k \leq 1 .
\end{equation}
In this case, the inner region $\{ 0 \leq r < r_k \}$ is described
by the closed RW metric with the scale factor $a_1(t)$ and the
outer region $\{ r_k < r \leq 1 \}$ by the open RW metric with
$a_2(t)$, satisfying $a_2(t) > a_1(t)$ as $t>0$. The result about
$q_D(t)$ for the case with above $k(r)$ and $\rho_0=3$ is
demonstrated in Fig.\ \ref{fig:ntc}.

As shown in Figs.\ \ref{fig:ntw} and \ref{fig:ntc}, for both
choices of $k(r)$, there exists the domain acceleration (at later
times). However, as to be shown in the following section, after
smoothing the step function invoked in above $k(r)$ in order to
get rid of the other singularity, we find no domain acceleration,
no matter how close to that in Eq.\
(\ref{eq:NT-k(r):L=1:exchange}) the function $k(r)$ is.

\subsection{Problems and Revision}

\begin{figure}[h]
  \centering
  \includegraphics[width=0.5\linewidth,clip]{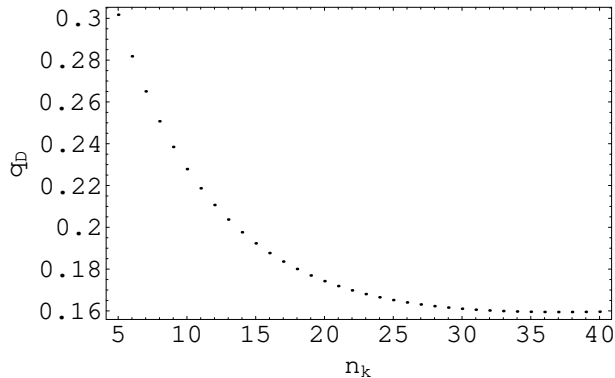}%
  \caption{The dependence of the deceleration parameter $q_D$ on
  the power $n_k$ in the function $k(r)$ in Eq.\ (\ref{eq:NT-smooth k(r)})
  for $t=1$, $\rho_0=3$ and $r_k=0.9$.}
  \label{fig:ntqton}
\end{figure}

In this section we use the following function $k(r)$ to replace
that involving a step function in the previous section.
\begin{equation} \label{eq:NT-smooth k(r)}
k(r)=-\frac{2 (r/r_k)^{n_k}}{1+(r/r_k)^{n_k}}+1 \, .
\end{equation}
When the power $n_k$ goes to infinity, the above function $k(r)$
approaches to that in Eq.\ (\ref{eq:NT-k(r):L=1:exchange})
involving a step function. If there is nothing wrong, we should
obtain the same result about $q_D(t)$ for the case invoking above
$k(r)$ with $n_k \rightarrow \infty$ and the case invoking $k(r)$
in Eq.\ (\ref{eq:NT-k(r):L=1:exchange}). For demonstration we
choose $t=1$, $\rho_0=3$ and $r_k=0.9$, corresponding to the case
where $q_D \sim -1$ as shown in Fig.\ \ref{fig:ntc} for $k(r)$ in
Eq.\ (\ref{eq:NT-k(r):L=1:exchange}). In Fig.\ \ref{fig:ntqton}
the dependence of $q_D$ on $n_k$ for the case with $k(r)$ in Eq.\
(\ref{eq:NT-smooth k(r)}) is illustrated. We can see that along
with the increasing of the power $n_k$ the domain deceleration
parameter $q_D$ always keeps far away from the value $-1$ (that
corresponds to significant acceleration), and smoothly approaches
to a positive value around $0.16$ (deceleration).

Where does this discrepancy come from? The discrepancy stems from
the doubtful calculation of the volume $V_D$ in
\cite{Nambu:2005zn} described in the previous section. In
\cite{Nambu:2005zn} Nambu and Tanimoto ignored the volume of the
junction (between two RW regions) at $r=r_k$ that is actually
significantly nonzero (even though $r=r_k$ corresponds to a ``2D
surface'' in the coordinate space) and should not be ignored.
After taking back the volume at $r=r_k$ into account as a more
appropriate treatment, we found no domain acceleration in the
cases discussed in \cite{Nambu:2005zn}.

\begin{figure}[h!]
  \centering
  \includegraphics[width=0.5\linewidth,clip]{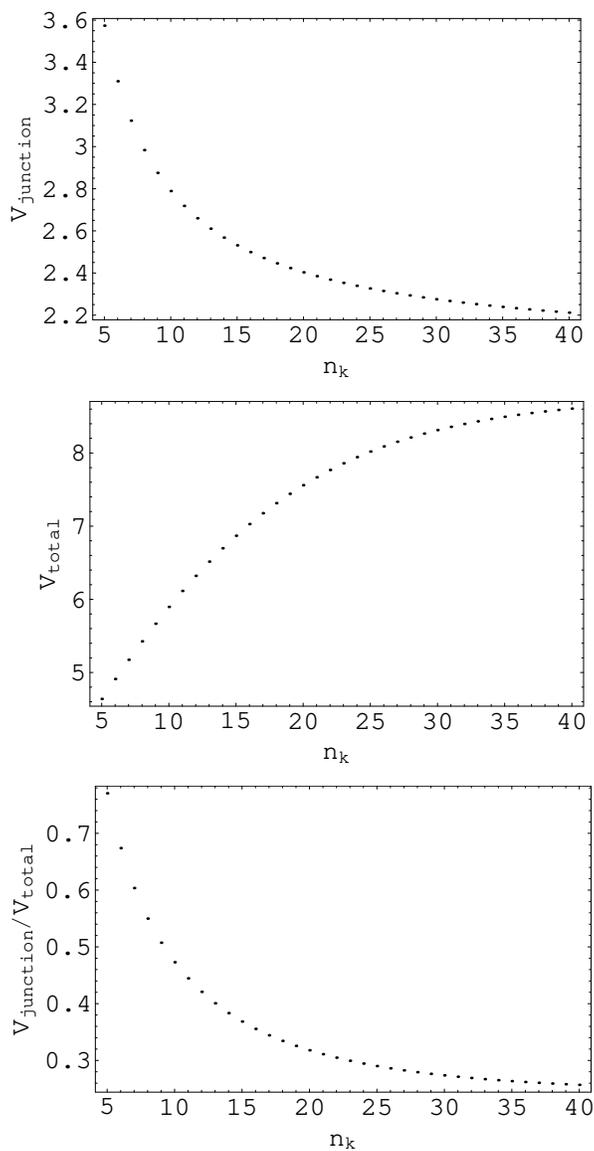}%
  \caption{The dependence of the volume of the junction and
  the total volume on the power $n_k$ in the function $k(r)$
  in Eq.\ (\ref{eq:NT-smooth k(r)}).}
  \label{fig:ntvton}
\end{figure}

In the following we study the volume of the junction region around
$r_k$ for the function $k(r)$ in Eq.\ (\ref{eq:NT-smooth k(r)}):
\begin{equation}\label{vjunc}
V_\textrm{junction} = 4 \pi
\int_{r_k-\varepsilon}^{r_k+\varepsilon}\frac{a^2 r^2
(a+a,_{r}r)dr}{\sqrt{1-k(r)r^2}} \, ,
\end{equation}
where
\begin{equation}
\varepsilon=\frac{1}{2 n_k} \, .
\end{equation}
The dependence of $V_\textrm{junction}$, $V_\textrm{total} \equiv
V_D + V_\textrm{junction}$ and
$V_\textrm{junction}/V_\textrm{total}$ on the power $n_k$ is
illustrated in Fig.\ \ref{fig:ntvton}. This figure shows that the
contribution to the total volume from the junction region remains
significantly nonzero when $n_k$ becomes large.

When $n_k$ goes to infinity and accordingly $\varepsilon$ goes to
zero, one can obtain a lower bound of $V_\textrm{junction}$ as
follows.
\begin{eqnarray} \label{eq:junction volume lower bound}
n_k \rightarrow \infty \, , \quad V_\textrm{junction}
&\rightarrow& 4 \pi \int_{r_k-\varepsilon}^{r_k+\varepsilon}
\frac{a^2 r^2 (a,_{r}r)dr}{\sqrt{1-k(r)r^2}} \nonumber \\
&>& \frac{4 \pi a_1^2 r_k^3}{\sqrt{1+r_k^2}}
\int_{r_k-\varepsilon}^{r_k+\varepsilon}a,_{r}dr \nonumber \\
&\cong& \frac{4 \pi a_1^2 r_k^3 (a_2-a_1)}{\sqrt{1+r_k^2}} \nonumber \\
&>& 0 \quad \textrm{ if } \; a_2 > a_1 \, ,
\end{eqnarray}
where $\varepsilon = (2n_k)^{-1}$ and
\begin{eqnarray}
a_1(t) &\equiv& \lim_{n_k \rightarrow \infty} a(t,r_k-\varepsilon) \, , \\
a_2(t) &\equiv& \lim_{n_k \rightarrow \infty} a(t,r_k+\varepsilon) \, . %
\end{eqnarray}
[Note that the first line of Eq.\ (\ref{eq:junction volume lower
bound}) is obtained by keeping only the singular part in the
integrand.] Thus, even when the junction region goes to a ``2D
surface'' in the coordinate space, its volume remains
significantly nonzero if the scale factor of the outer open
under-density region ($a_2$) is significantly larger than that of
the inner closed over-density region ($a_1$). As mentioned in the
previous section, with the initial condition in Eq.\
(\ref{eq:NT-initial condition}) we do have $a_2 > a_1$ for $t>0$.

Note that by changing the initial condition for $a_1(t)$ and
$a_2(t)$ [instead of using that in Eq.\ (\ref{eq:NT-initial
condition})] Nambu and Tanimoto's treatment could become valid
during some specific time period when $a_1 \simeq a_2$, which
might be achieved through, for example, choosing non-trivial
$t_b(r)$ (i.e., making the big bang or the expansion at different
places begin at different times). This may be the reason why
non-trivial $t_b(r)$ is a necessary ingredient in our search for
domain acceleration.


\end{document}